\documentclass[11pt,amsmath,nofootinbib,superscriptaddress]{revtex4}
\usepackage{amssymb}
\usepackage[hypertex]{hyperref}
\usepackage{graphicx}
\usepackage{epstopdf}
\usepackage{bm}
\usepackage{epsfig}
\usepackage{graphics}
\usepackage{xspace}
\usepackage{amsfonts}
\usepackage{color}

\begin{document}


\title{Signature of the $\gamma$+jet and dijet production mediated by an excited quark with QCD next-to-leading order accuracy at the LHC}

\vspace*{1cm}

\author{}

\author{Yong Chuan Zhan}
\affiliation{School of Physics and State Key Laboratory of Nuclear Physics and Technology, Peking University, Beijing, 100871, China}

\author{Chong Sheng Li\footnote{csli@pku.edu.cn}}
\affiliation{School of Physics and State Key Laboratory of Nuclear Physics and Technology, Peking University, Beijing, 100871, China}
\affiliation{Center for High Energy Physics, Peking University, Beijing, 100871, China}

\author{Ze Long Liu}
\affiliation{School of Physics and State Key Laboratory of Nuclear Physics and Technology, Peking University, Beijing, 100871, China}

\author{Shi Ang Li}
\affiliation{School of Physics and State Key Laboratory of Nuclear Physics and Technology, Peking University, Beijing, 100871, China}



\begin{abstract}
 \vspace*{0.3cm}
We present a detailed study of the production and decay of the excited quark at the QCD next-to-leading order (NLO) level at the Large Hadron Collider, using the narrow width approximation and helicity amplitudes method. We find that the QCD NLO corrections can tighten the constraints on the model parameters and reduce the scale dependencies of the total cross sections. We discuss the signals of the excited quark production with decay mode $q^{\ast}\rightarrow q\gamma$ and $q^{\ast}\rightarrow qg$, and present several important kinematic distributions. Moreover, we give the upper limits of the excited quark excluded mass range and the allowed parameter space for the coupling constants and the excited quark mass.
\end{abstract}
\maketitle
\newpage



\section{Introduction}
\label{sec:intro}

Although the standard model (SM) successfully describes a wide range of phenomena in particle physics, there are still many questions left
unanswered. The SM is still regarded as an effective theory, a low-energy approximation of a more fundamental theory.
There are three popular types of new physics theories: (i) models with an extended (family) symmetry or scalar sector, (ii) higher dimensional theory, and (iii) quark-lepton compositeness (namely, the SM fermions are not elementary anymore \cite{Bhattacharya:2009xg}). Here we are mainly concerned with the third type, where
the substructure of quarks and leptons will lead to a rich spectrum of new particles with unusual quantum numbers such as the excited quarks \cite{Cakir:1999nu}.
In the SM there is not a resonance production which decays into a photon+jet pair in the proton-proton ($pp$) collisions, and direct photon+jet production at tree level occurs via scattering of a quark and a gluon or through quark-antiquark annihilation, where the photon+jet invariant mass ($m_{\gamma j}$) distribution is rapidly falling; thus, the photon+jet production mediated by a heavy excited quark may be discovered if it exists.
Besides, the search for the dijet production mediated by the excited quark is also interesting for the experiment at the LHC.

In general, the interactions between the excited quarks ($q^{\ast}$) and gauge bosons can be written as
\begin{equation}
\mathcal{L}_{\text{gauge}}=\overline{q^{\ast}}\gamma^{\mu}\left[g_s\frac{\lambda_a}{2}G_{\mu}^a + g\frac{\tau}{2}W_{\mu}
 + g^{\prime}\frac{Y}{2}B_{\mu}\right]q^{\ast}.
\end{equation}
Gauge bosons can also induce transitions between ordinary (left-handed) and excited (right-handed) quarks. Such interactions can be described by the effective Lagrangian as follows \cite{Baur:1987ga,Baur:1989kv,Bhattacharya:2009xg}:
\begin{equation}
\mathcal{L}_{\text{trans}}=\frac{1}{2\Lambda}\overline{q^{\ast}_R}\sigma^{\mu\nu}
\left[g_s f_s \frac{\lambda_a}{2}G_{\mu\nu}^a + g f \frac{\tau}{2}W_{\mu\nu} + g^{\prime}f^{\prime}\frac{Y}{2}B_{\mu\nu}\right]
q_L + H.c.,
\end{equation}
where $G_{\mu\nu}^a$, $W_{\mu\nu}$ and $B_{\mu\nu}$ are the field strength tensors of the SU(3), SU(2) and U(1) gauge fields, respectively. Here, $g_s$, $g=e/\sin\theta_{W}$, $g^{\prime}=e/\cos\theta_{W}$ are the strong and electroweak gauge couplings, $\lambda_a$ is the Gell-Mann matrix, $\tau$ is the Pauli matrix, and the weak hypercharge is $Y=1/3$, respectively. $\Lambda$ is the typical compositeness energy scale, and $f_s$, $f$, $f^{\prime}$ are parameters determined by composite dynamics, which represent the strength of the interactions between the excited quarks and their SM partners.

In $pp$ collisions, the excited quarks can decay into a quark and a gauge boson ($\gamma$, $g$, $W$, $Z$), and the resonance production would be observed in the invariant mass distribution of the decay products.
These processes have been widely discussed \cite{Baur:1987ga,Baur:1989kv,Cakir:1999nu,Cakir:2000sw,Cakir:2000vt,Bhattacharya:2009xg}, where the production of the excited quarks decaying into a photon+jet, dijet and W/Z+jet at the leading-order (LO) level is discussed in Refs.\cite{Bhattacharya:2009xg,Cakir:1999nu} and Refs.\cite{Cakir:2000sw,Cakir:2000vt}, respectively.
Many searches of the signals of the excited quark at the LHC have been performed in various decay channels \cite{ATLAS:2011ai,Aad:2013cva,Khachatryan:2014aka, Chatrchyan:2013qha,Khachatryan:2015sja,Aad:2014aqa,
Chatrchyan:2012tw,Aad:2013rna}.
But the LO predictions suffer from large scale uncertainties, and cannot match the expected experimental accuracy at the LHC.
In this paper, we investigate the signature of the photon+jet and dijet production mediated by the excited quark with the QCD next-to-leading order (NLO) accuracy at the LHC.

The arrangement of this paper is as follows. In Sec.\ref{sec:NWA.Hel}, we briefly describe the narrow width approximation and helicity amplitudes method used in our calculation. In Sec.\ref{sec:LO}, we show the LO results for these processes. In Sec.\ref{sec:NLO}, we present the details of the QCD NLO calculations for the production and decay processes. In Sec.\ref{sec:numerical}, we investigate the numerical results, where we discuss the scale uncertainties and give some important kinematic distributions. In Sec.\ref{sec:simulation}, we discuss the signal and background of the excited quark. A conclusion will be given in Sec.\ref{sec:conclusion}.

\section{Narrow Width Approximation And Helicity Amplitudes Method}
\label{sec:NWA.Hel}

The narrow width approximation \cite{Kauer:2007zc,Uhlemann:2008pm} is often used for a resonant process when the heavy resonance has a small decay width, which greatly simplifies the calculation. For a scalar resonance, the total cross section can be written as
\begin{eqnarray}
\sigma &=& \frac{(2\pi)^7}{2s}\int_{q^2_\text{min}}^{q^2_\text{max}}dq^2\int d\phi_p d\phi_d
|\mathcal{M} _p(q^2)|^2 \left[\left(q^2-m^2\right)^2+(m\Gamma)^2\right]^{-1} |\mathcal{M} _d(q^2)|^2
               \nonumber \\
 &=& \frac{(2\pi)^8}{4s m\Gamma}\int d\phi_p|\mathcal{M} _p(m^2)|^2 \int d\phi_d|\mathcal{M} _d(m^2)|^2,
\end{eqnarray}
where $m$ denotes the mass of the resonance and $\Gamma$ denotes its decay width. $\mathcal{M}_p$ and $\mathcal{M}_d$ are the amplitudes of the production and decay parts, respectively. In our case, the resonance is an excited quark, and the intermediate excited quark is on shell, so its propagator can be expressed as
\begin{equation}
\frac{\slash\!\!\!q+m}{q^2-m^2+im\Gamma}=\frac{u_+\bar{u}_+ + u_-\bar{u}_-}{q^2-m^2+im\Gamma},
\end{equation}
where $u_+$ and $u_-$ denote the quark spinors with positive and negative helicities, respectively. Thus the cross section can be written as
\begin{eqnarray}
\sigma &=&
\frac{(2\pi)^8}{4s m\Gamma}\bigg(\int d\phi_p |\mathcal{M} _p^+|^2 \int d\phi_d|\mathcal{M} _d^+|^2
+\int d\phi_p |\mathcal{M} _p^-|^2  \int d \phi_d  |\mathcal{M} _d^-|^2\bigg),
\end{eqnarray}
where $\mathcal{M}_p^{\pm}$ and $\mathcal{M}_d^{\pm}$ are the helicity amplitudes of the production and decay, respectively.

Here we have  adopted the modified helicity amplitude method, which is suitable for massive particles \cite{Kleiss:1985yh,Badger:2010mg}, in our calculations, where the massless and massive spinors are defined as
\begin{equation}
|k^{\pm}\rangle=u_{\pm}(k)=v_{\mp}(k), \qquad \langle k^{\pm}|=\bar{u}_{\pm}(k)=\bar{v}_{\mp}(k),
\end{equation}
and
\begin{eqnarray}
\nonumber
u_{\pm}(p,m;\eta,p^{\flat}) = \frac{(\slash\!\!\! p+m)|\eta^{\mp}\rangle}{\langle p^{\flat\pm}|\eta^{\mp}\rangle}, &&
\bar{u}_{\pm}(p,m;\eta,p^{\flat}) = \frac{\langle\eta^{\mp}|(\slash\!\!\! p+m)}{\langle \eta^{\mp}|p^{\flat\pm}\rangle}, \\
v_{\pm}(p,m;\eta,p^{\flat}) = \frac{(\slash\!\!\! p-m)|\eta^{\pm}\rangle}{\langle p^{\flat\mp}|\eta^{\pm}\rangle}, &&
\bar{v}_{\pm}(p,m;\eta,p^{\flat}) = \frac{\langle\eta^{\pm}|(\slash\!\!\! p-m)}{\langle \eta^{\pm}|p^{\flat\mp}\rangle},
\end{eqnarray}
respectively. Here $p^{\flat}$ and $\eta$ are auxiliary massless momenta, which satisfy
\begin{equation}
p=p^{\flat}+\frac{m^2}{2p\cdot\eta}\eta, \qquad p^2=m^2, \qquad (p^{\flat})^2=\eta^2=0.
\end{equation}

\section{Leading-Order Results}
\label{sec:LO}

\begin{figure}[h]
  \centering
  \includegraphics[scale=1]{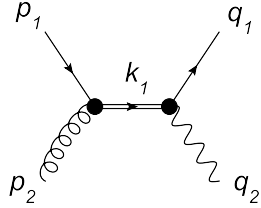}\\
  \caption{The LO Feynman diagram for the production and decay of the excited quark in the narrow width approximation.}
  \label{Fig. FD_born222}
\end{figure}

In proton-proton collisions, the excited quarks can be produced via quark-gluon fusion $qg\rightarrow q^{\ast}$, and then decay into a quark and a gauge boson ($\gamma$, $g$, $W$, $Z$) at the LO. In the narrow width approximation, the corresponding Feynman diagram is shown in Fig.[\ref{Fig. FD_born222}].

The LO helicity amplitudes for the production and decay of the excited quark are
\begin{eqnarray}
\nonumber
\mathcal{M}_{qg\rightarrow q^{\ast}}^{-++} &=& -\frac{i \sqrt{2}\ g_s f_s t^a_{ji}\ [p_1|p_2] [p_2|k_1|\eta(k_1)\rangle }
{\Lambda \ \langle p^{\flat}(k_1)|\eta(k_1)\rangle }, \\
\nonumber
\mathcal{M}_{qg\rightarrow q^{\ast}}^{-+-} &=& \frac{i \sqrt{2}\ g_s f_s m_{q^{\ast}} t^a_{ji}\ [p_1|p_2] [p_2|\eta(k_1)] }
{\Lambda \ [p^{\flat}(k_1)|\eta(k_1)] }, \\
\nonumber
\mathcal{M}_{q^{\ast}\rightarrow q\gamma}^{+++} &=& \frac{i \sqrt{2}\ e f_{\gamma} m_{q^{\ast}} \ [q_1|q_2] [q_2|\eta(k_1)] }
{\Lambda \ [p^{\flat}(k_1)|\eta(k_1)] }, \\
\nonumber
\mathcal{M}_{q^{\ast}\rightarrow q\gamma}^{-++} &=& \frac{i \sqrt{2}\ e f_{\gamma} \ [q_1|q_2] [q_2|k_1|\eta(k_1)\rangle }
{\Lambda \ \langle p^{\flat}(k_1)|\eta(k_1)\rangle }, \\
\nonumber
\mathcal{M}_{q^{\ast}\rightarrow qg}^{+++} &=& \frac{i \sqrt{2}\ g_s f_s m_{q^{\ast}} t^b_{kj}\ [q_1|q_2] [q_2|\eta(k_1)] }
{\Lambda \ [p^{\flat}(k_1)|\eta(k_1)] }, \\
\mathcal{M}_{q^{\ast}\rightarrow qg}^{-++} &=& \frac{i \sqrt{2}\ g_s f_s t^b_{kj}\ [q_1|q_2] [q_2|k_1|\eta(k_1)\rangle }
{\Lambda \ \langle p^{\flat}(k_1)|\eta(k_1)\rangle }.
\end{eqnarray}
The LO decay widths of the excited quark are given by
\begin{eqnarray}\label{Eq. LO decay width}
\nonumber
\Gamma(q^{\ast}\rightarrow qg) &=& \frac{1}{3}\alpha_s f_s^2 \frac{m_{q^{\ast}}^3}{\Lambda^2}, \\
\nonumber
\Gamma(q^{\ast}\rightarrow q\gamma) &=& \frac{1}{4}\alpha f_{\gamma}^2 \frac{m_{q^{\ast}}^3}{\Lambda^2}, \\
\Gamma(q^{\ast}\rightarrow qV) &=& \frac{1}{32\pi}g_V^2 f_V^2 \frac{m_{q^{\ast}}^3}{\Lambda^2}
\left(1-\frac{m_V^2}{m_{q^{\ast}}^2}\right)^2\left(2+\frac{m_V^2}{m_{q^{\ast}}^2}\right),
\end{eqnarray}
with
\begin{eqnarray}
\nonumber
&& f_{\gamma}=f\tau_3+f^{\prime}Y/2, \quad f_W=f/\sqrt{2}, \quad f_Z=f\tau_3\cos^2\theta_W-f^{\prime}(Y/2)\sin^2\theta_W, \\
&& g_W=e/\sin\theta_W, \quad g_Z=g_W/\cos\theta_W,
\end{eqnarray}
where $V$ represents $W$ or $Z$, and $\tau_3$ denotes the third component of the weak isospin of the excited quark.

The total cross section at hadron colliders can be obtained by convoluting the partonic cross section with the parton distribution functions (PDFs) $G_{i/P}$:
\begin{equation}
\sigma_{pp\rightarrow q^{\ast}\rightarrow qX}=\sum_{ij}\int dx_1dx_2 G_{i/P_1}(x_1,\mu_f)G_{j/P_2}(x_2,\mu_f)\sum_{k=+,-}\hat{\sigma}^k_{ij}
\frac{\Gamma^k_{q^{\ast}\rightarrow qX}}{\Gamma},
\end{equation}
where $k$ denotes the helicity of the excited quark, $\Gamma$ is the total decay width of the excited quark, and $\mu_f$ is the factorization scale.

\section{QCD NLO Corrections}
\label{sec:NLO}

\begin{figure}[h]
  \centering
  \includegraphics[scale=1]{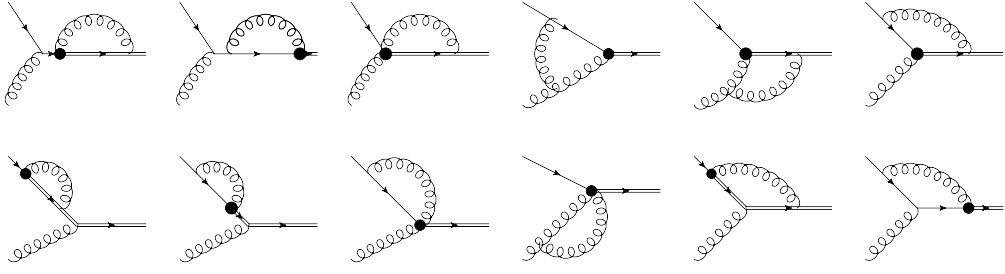}\\
  \caption{Feynman diagrams for the virtual corrections to the production of the excited quark. The virtual corrections for the $q^{\ast}\rightarrow qg$ decay are the same.}
  \label{Fig. FD_pro_virt}
\end{figure}

\begin{figure}[h]
  \centering
  \includegraphics[scale=1]{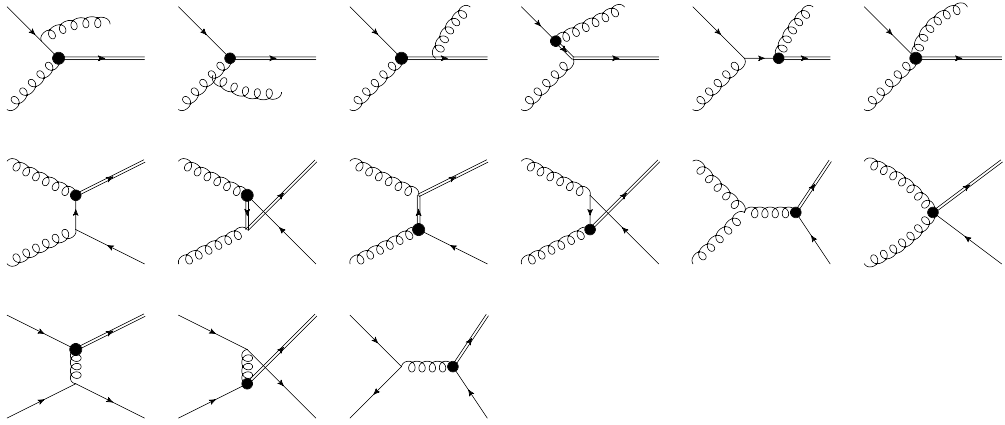}\\
  \caption{Feynman diagrams for the real corrections to the production of the excited quark.}
  \label{Fig. FD_pro_real}
\end{figure}

\begin{figure}[h]
  \centering
  \includegraphics[scale=1]{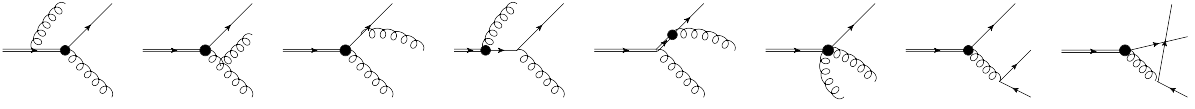}\\
  \caption{Feynman diagrams for the real corrections to the $qg$ decay channel of the excited quark.}
  \label{Fig. FD_decay_qg}
\end{figure}

\begin{figure}[h]
  \centering
  \includegraphics[scale=1]{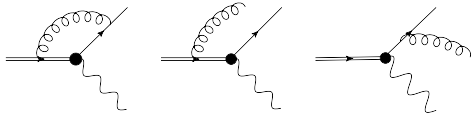}\\
  \caption{Feynman diagrams for the virtual and real corrections to the $q\gamma/qZ/qW$ decay channel of the excited quark.}
  \label{Fig. FD_decay_qga}
\end{figure}

The Feynman diagrams for the QCD NLO corrections to the production and decay of the excited quark are shown in Figs.
\ref{Fig. FD_pro_virt}-\ref{Fig. FD_decay_qga}.
The interference between the production and decay at the NLO level are neglected, since their contributions are suppressed by $\mathcal{O}(\frac{\Gamma}{m})$ \cite{Fadin:1993dz,Fadin:1993kt,Melnikov:1993np}.
We also compared the LO differential distributions in the narrow width approximation with the ones without such an approximation. There are only slight changes in the shapes, so using the narrow width approximation is reasonable in our scenario.
We use the 't Hooft-Veltman(HV) scheme \cite{'tHooft:1972fi,Kilgore:2011ta,Bern:2002zk} in $n=4-2\epsilon$ dimensions to regularize the ultraviolet (UV), soft infrared and collinear divergences in the virtual loop corrections.
For the real emissions, the two cutoff phase-space slicing method \cite{Harris:2001sx} is used to separate the infrared singularities.


\subsection{Virtual corrections}
The virtual corrections contain both UV and infrared (IR) divergences, and the UV divergences can be canceled by introducing counterterms. All of the renormalization constants are given by
\begin{eqnarray}
\nonumber
\delta Z_2^{(q)} &=& -\frac{\alpha_s}{3\pi}C_{\epsilon}\left(\frac{1}{\epsilon_{\text{UV}}}-\frac{1}{\epsilon_{\text{IR}}}\right), \\
\nonumber
\delta Z_2^{(g)} &=&
  -\frac{\alpha_s}{2\pi}C_{\epsilon}\left(\frac{N_f}{3}-\frac{5}{2}\right)
  \left(\frac{1}{\epsilon_{\text{UV}}}-\frac{1}{\epsilon_{\text{IR}}}\right)
  -\frac{\alpha_s}{6\pi}C_{\epsilon}\frac{1}{\epsilon_{\text{UV}}}, \\
\nonumber
\delta Z_{g_s} &=&
  \frac{\alpha_s}{4\pi}C_{\epsilon}\left(\frac{N_f}{3}-\frac{11}{2}\right)\frac{1}{\epsilon_{\text{UV}}}
  +\frac{\alpha_s}{12\pi}C_{\epsilon}\frac{1}{\epsilon_{\text{UV}}}, \\
\nonumber
\delta Z_2^{(q^{\ast})} &=& -\frac{\alpha_s}{3\pi}C_{\epsilon}
  \left(\frac{1}{\epsilon_{\text{UV}}}+\frac{2}{\epsilon_{\text{IR}}}+4+3\ln\frac{\mu_r^2}{m_{q^{\ast}}^2}\right), \\
\nonumber
\delta Z_{f_s} &=& \frac{\alpha_s}{6\pi}C_{\epsilon}\frac{1}{\epsilon_{\text{UV}}}, \\
\delta Z_{f_{\gamma}} &=& \frac{\alpha_s}{3\pi}C_{\epsilon}\frac{1}{\epsilon_{\text{UV}}},
\end{eqnarray}
with
\begin{equation}
C_{\epsilon} = \frac{(4\pi)^{\epsilon}}{\Gamma(1-\epsilon)}.
\end{equation}
Here, for the external fields and the coupling constants, we fix the relevant renormalization constants using on-shell subtraction and the $\overline{\text{MS}}$ scheme, respectively. After renormalization, the results of the virtual corrections for the production of the excited quark still contain the infrared divergences
\begin{eqnarray}\label{Eq. virt pro}
\nonumber
\mathcal{M}_{qg\rightarrow q^{\ast}}^{\text{virt}} &=& \mathcal{M}_{qg\rightarrow q^{\ast}}^{\text{born}}
  \times\frac{\alpha_s}{12\pi}C_{\epsilon}\bigg[
  -\frac{13}{\epsilon_{\text{IR}}^2}+\frac{1}{\epsilon_{\text{IR}}}\left(N_f-\frac{53}{2}
  -13\ln(\mu_r^2/m_{q^{\ast}}^2)\right) \\
\nonumber
&&  -23+\frac{23\pi^2}{6}-12\ln(\mu_r^2/m_{q^{\ast}}^2)
  -\frac{13}{2}\ln^2(\mu_r^2/m_{q^{\ast}}^2) \\
&&  +2m_{q^{\ast}}^2C_0(0,0,m_{q^{\ast}}^2,0,m_{q^{\ast}}^2,m_{q^{\ast}}^2)\bigg],
\end{eqnarray}
where $C_0$ is a finite scalar integral defined as \cite{Ellis:2007qk}
\begin{eqnarray}
\nonumber
&& C_0(p_1^2,p_2^2,p_3^2;m_1^2,m_2^2,m_3^2) = \\
&& \qquad \frac{\mu^{4-d}}{i\pi^{d/2}r_{\Gamma}}\int d^d l
\frac{1}{(l^2-m_1^2+i\epsilon)((l+p_1)^2-m_2^2+i\epsilon)((l+p_1+p_2)^2-m_3^2+i\epsilon)},
\end{eqnarray}
with
\begin{equation}
r_{\Gamma}=\frac{\Gamma^2(1-\epsilon)\Gamma(1+\epsilon)}{\Gamma(1-2\epsilon)}.
\end{equation}
Here the infrared divergences include the soft infrared divergences and the collinear infrared divergences. The soft infrared divergences can be canceled by adding the real corrections, and the remaining collinear infrared divergences can be absorbed in the redefinition of the PDF \cite{Altarelli:1979ub}, which will be discussed in the following subsections.

\subsection{Real corrections for the production of the excited quark}
\subsubsection{Real gluon emission}
For the real gluon emission, we adopt the two cutoff phase-space slicing method to isolate the soft and collinear singularities, which introduces two small cutoff parameters $\delta_s$ and $\delta_c$ to separate the phase space into three regions: soft, hard collinear and hard noncollinear. The hard-noncollinear part is finite and can be calculated numerically.

In the soft region $E_s\leq\delta_s\sqrt{s}/2$, the matrix element factorizes as \cite{Harris:2001sx}
\begin{equation}
\mathcal{M}_{\text{soft}}\simeq g_s\mu_r^{\epsilon}\epsilon^{\mu}(p_s)J_{\mu}^a(p_s)\mathcal{M}_{\text{born}},
\end{equation}
where the non-Abelian eikonal current is given by
\begin{equation}
J_{\mu}^a(p_s)=\sum_{f}T^a_f\frac{p_f^{\mu}}{p_f\cdot p_s}.
\end{equation}
Here if the soft gluon is emitted from a final-state quark or initial-state anti-quark, the color charge is $T^a_{ij}=-t^a_{ij}$, while for a final-state anti-quark or initial-state quark, it is $T^a_{ij}=t^a_{ij}$. If the emitting parton is a gluon, the color charge is $T^a_{bc}=if_{abc}$. The cross section in the soft region can be written as
\begin{equation}
d\hat{\sigma}_{\text{soft}}=\bigg[\frac{\alpha_s}{2\pi}\frac{\Gamma(1-\epsilon)}{\Gamma(1-2\epsilon)}\bigg(\frac{4\pi\mu_r^2}{s}\bigg)
^{\epsilon}\bigg]\sum_{f,f^{\prime}}d\hat{\sigma}_{\text{born}}^{f,f^{\prime}}\int\frac{-p_f\cdot p_{f^{\prime}}}
{p_f\cdot p_s p_{f^{\prime}}\cdot p_s}dS,
\end{equation}
with
\begin{equation}
dS=\frac{1}{\pi}\bigg(\frac{4}{s}\bigg)^{-\epsilon}\int_0^{\delta_s\sqrt{s}/2}dE_sE_s^{1-2\epsilon}
\sin^{1-2\epsilon}\theta_1d\theta_1\sin^{-2\epsilon}\theta_2d\theta_2.
\end{equation}
Thus we obtain the result for the soft radiation of the production of the excited quark
\begin{eqnarray}\label{Eq. soft pro}
\nonumber
d\hat{\sigma}_{qg\rightarrow q^{\ast}+g}^{\text{soft}} &=& d\hat{\sigma}_{qg\rightarrow q^{\ast}}^{\text{born}}\frac{\alpha_s}{3\pi}C_{\epsilon}\bigg[\frac{13}{2\epsilon_{\text{IR}}^2}
+\frac{1}{\epsilon_{\text{IR}}}\bigg(2+\frac{13}{2}\ln\frac{\mu_r^2}{m_{q^{\ast}}^2}-13\ln\delta_s\bigg) \\
&& +4-\frac{13\pi^2}{12}+13\ln^2\delta_s -4\ln\delta_s+2\ln\frac{\mu_r^2}{m_{q^{\ast}}^2}+\frac{13}{4}\ln^2\frac{\mu_r^2}{m_{q^{\ast}}^2} -13\ln\delta_s\ln\frac{\mu_r^2}{m_{q^{\ast}}^2}\bigg].
\end{eqnarray}

In the hard-collinear region, $E_s\geq\delta_s\sqrt{s}/2$ and the invariants ($s_{ij}$ or $t_{ij}$) are smaller in magnitude than $\delta_c s$;
 the amplitude squared can be factorized into~\cite{Collins:1985ue,Bodwin:1984hc}
\begin{equation}
\overline{\sum}|\mathcal {M}_{qg \rightarrow q^{\ast}+g}^{\text{coll}}|^2 {\rightarrow} (4\pi\alpha_s\mu^{2\epsilon}_r) \overline{\sum}|\mathcal {M}_{qg \rightarrow q^{\ast}}^{\text{born}}|^2 \left[\frac{-2P_{gg}(z,\epsilon)}{zt_{24}}
+\frac{-2P_{qq}(z,\epsilon)}{zt_{14}}\right],
\end{equation}
where $z$ denotes the fraction of the momentum carried by $q(g)$, and the unregulated Altarelli-Parisi splitting functions are given by~\cite{Harris:2001sx}
 \begin{eqnarray}
P_{qq}(z,\epsilon) &=& C_{F}\Bigg[\frac{1+z^{2}}{1-z}-\epsilon(1-z)\Bigg], \nonumber \\
P_{gg}(z,\epsilon) &=& 2C_A\Bigg[\frac{z}{1-z}+\frac{1-z}{z}+z(1-z)\Bigg].
\end{eqnarray}
Then the phase space in the collinear limit $-\delta_c s < t_{i4} < 0$ can be factorized as
\begin{equation}
d\text{PS}(qg \rightarrow q^{\ast} + g)\stackrel{\text{coll}}{\longrightarrow} d\text{PS}(qg \rightarrow q^{\ast}; s^{\prime} = zs)
\frac{(4\pi)^{\epsilon}}{16\pi^2\Gamma(1-\epsilon)}dzdt_{i4}[-(1-z)t_{i4}]^{-\epsilon}.
\end{equation}
Therefore, after convoluting with the PDFs, the cross section
in the hard-collinear region can be written as~\cite{Harris:2001sx}
\begin{eqnarray}
\sigma^{HC} & = & \int dx_1dx_2 ~\hat{\sigma}^B_{qg}
\left[\frac{\alpha_s}{2\pi}\frac{\Gamma(1-\epsilon)}{\Gamma(1-2\epsilon)}
\left(\frac{4\pi\mu^2_r}{s}\right)^{\epsilon}\right]\left(-\frac{1}{\epsilon}\right)
\delta_c^{-\epsilon}\bigg[P_{qq}(z,\epsilon)G_{q/p}(x_1/z)G_{g/p}(x_2)
\nonumber\\&& +
P_{gg}(z,\epsilon)G_{g/p}(x_2/z)G_{q/p}(x_1) +
(x_1 \leftrightarrow x_2)\bigg]
\frac{dz}{z}\left(\frac{1-z}{z}\right)^{-\epsilon},
\end{eqnarray}
where $G_{q(g)/p}(x)$ is the bare PDF.

\subsubsection{Massless (anti)quark emission}
In addition to the real gluon emission, an additional massless
$q(\bar q)$ in the final state should be taken into consideration. These contributions contain initial-state collinear singularities, which can be isolated using the two cutoff phase-space slicing method. And the
cross section for the process with an additional massless
$q(\bar{q})$ emission, including the noncollinear part and the collinear part, can be written as \cite{Harris:2001sx}
\begin{eqnarray}
 \sigma^{add} & = & \int dx_1dx_2\sum_{(\alpha=q,\bar{q},q')} \Bigg\{
\hat{\sigma}^{\overline{C}}\left(q \alpha \rightarrow q^{\ast} + q(\bar{q})\right)G_{q/p}(x_1)G_{\alpha/p}(x_2)+
\hat{\sigma}^B_{qg} \left[\frac{\alpha_s}{2\pi}\frac{\Gamma(1-\epsilon)}{\Gamma(1-2\epsilon)} \left(\frac{4\pi\mu^2_r}{s}\right)^{\epsilon}\right] \nonumber\\
&& \times \left(-\frac{1}{\epsilon}\right) \delta_c^{-\epsilon}
P_{g\alpha}(z,\epsilon)G_{q/p}(x_1)G_{\alpha/p}(x_2/z)\frac{dz}{z}\left(\frac{1-z}{z}\right)^{-\epsilon}+ (x_1\leftrightarrow x_2)\Bigg\}  \nonumber\\
&& + \int dx_1dx_2 ~\Bigg\{\hat{\sigma}^{\overline{C}} (gg \rightarrow q^{\ast} +\bar{q})G_{g/p}(x_1)G_{g/p}(x_2)+
\hat{\sigma}^B_{qg} \left[\frac{\alpha_s}{2\pi}\frac{\Gamma(1-\epsilon)}{\Gamma(1-2\epsilon)} \left(\frac{4\pi\mu^2_r}{s}\right)^{\epsilon}\right] \nonumber\\
&& \times \left(-\frac{1}{\epsilon}\right) \delta_c^{-\epsilon} P_{qg}(z,\epsilon)G_{g/p}(x_1/z)G_{g/p}(x_2) \frac{dz}{z}\left(\frac{1-z}{z}\right)^{-\epsilon}+ (x_1 \leftrightarrow x_2)\Bigg\},
\end{eqnarray}
with
\begin{eqnarray}
P_{qg}(z,\epsilon) & = & P_{\bar{q}g}(z) =
\frac{1}{2}\left[z^2+(1-z)^2\right]-z(1-z)\epsilon,\nonumber\\
P_{gq}(z,\epsilon) & = & P_{g\bar{q}}(z) = C_F\left[\frac{z}{1+(1-z)^2}-z\epsilon\right],
\end{eqnarray}
where $\hat{\sigma}^{\overline{C}}$ are the noncollinear cross sections for the processes of the $q\bar{q}(q,q')$, $gg$ and $q'\bar{q'}$ initial states,
\begin{equation}
\hat{\sigma}^{\overline{C}} = \frac{1}{2s}\int d \text{PS}_{\overline{C}} \Big\{\sum_{(\alpha=q,\bar{q},q')}
|\mathcal {M}(q\alpha\rightarrow q^{\ast}+\alpha)|^2+ |\mathcal {M}(gg \rightarrow q^{\ast}+\bar{q})|^2 +
 |\mathcal {M}(q'\bar{q'}\rightarrow q^{\ast}+\bar{q})|^2\Big\}.
\end{equation}

\subsubsection{Mass factorization}
The soft divergences can be canceled out after adding the renormalized virtual corrections and the real corrections together. But there still remain collinear divergences which can be factorized into a redefinition of the PDFs.
In the $\overline{\text{MS}}$ convention, the scale-dependent PDF
$G_{\alpha/p}(x,\mu_f)$ can be written as ~\cite{Harris:2001sx}
\begin{equation}
 G_{\alpha/p}(x,\mu_f)  =  G_{\alpha/p}(x) +
\sum_{\beta}\left(-\frac{1}{\epsilon}\right)\left[
\frac{\alpha_s}{2\pi}\frac{\Gamma(1-\epsilon)}{\Gamma(1-2\epsilon)}
 \left(\frac{4\pi\mu^2_r}{\mu_f^2}\right)^{\epsilon}\right]
 \int_x^1 \frac{dz}{z} P_{\alpha\beta}(z) G_{\beta/p}(x/z),
\end{equation}
where the Altarelli-Parisi splitting function is defined as
\begin{eqnarray}
P_{\alpha\beta}(y,\epsilon) = P_{\alpha\beta}(y) +\epsilon P'_{\alpha\beta}(y).
\end{eqnarray}
The resulting $\mathcal{O} (\alpha_s)$ expression for the initial-state collinear contribution
is
\begin{eqnarray}
 \sigma^{coll} &=&  \int dx_1 dx_2 \hat{\sigma}^B_{qg}\bigg[\frac{\alpha_s}{2\pi}
\frac{\Gamma(1-\epsilon)} {\Gamma(1-2\epsilon)}
\bigg(\frac{4\pi\mu^2_r}{s}\bigg)^\epsilon \bigg] \nonumber\\
&& \times \bigg\{\tilde{G}_{q/p}(x_1,\mu_f) G_{g/p}(x_2,\mu_f) +
G_{q/p}(x_1,\mu_f) \tilde{G}_{g/p}(x_2,\mu_f)+ \nonumber
\\ &&
\sum_{\alpha=q,g}\bigg[\frac{A_1^{sc}(\alpha\rightarrow
\alpha g)}{\epsilon} +A_0^{sc}(\alpha\rightarrow \alpha
g)\bigg]G_{q/p}(x_1,\mu_f) G_{g/p}(x_2,\mu_f)+
(x_1\leftrightarrow x_2)\bigg\} ,
\end{eqnarray}
where
\begin{eqnarray}
A_1^{sc}(q\rightarrow qg)&=&C_F\left(2\ln\delta_s +\frac{3}{2}\right), \nonumber\\
A_1^{sc}(g\rightarrow gg)&=&2C_A\ln\delta_s + \frac{11C_A-2N_f}{6}, \nonumber\\
A_0^{sc}&=&A_1^{sc}\ln\left(\frac{s}{\mu_f^2}\right),
\end{eqnarray}
and
\begin{eqnarray}
\tilde{G}_{\alpha/p}(x,\mu_f)&=&\sum_{\beta}\int_x^{1-
\delta_s\delta_{\alpha\beta}} \frac{dy}{y}
G_{\beta/p}\left(\frac{x}{y},\mu_f\right)\tilde{P}_{\alpha\beta}(y),
\end{eqnarray}
with
\begin{eqnarray}
\tilde{P}_{\alpha\beta}(y)=P_{\alpha\beta}(y) \ln\left(\delta_c
\frac{1-y}{y} \frac{s}{\mu_f^2}\right) -P_{\alpha\beta}'(y).
\end{eqnarray}

Finally, the NLO total cross section for $pp\rightarrow q^{\ast}$ in the $\overline{\text{MS}}$ factorization scheme is
\begin{eqnarray}
\nonumber
\sigma^{\text{NLO}} &=& \int\{dx_1dx_2[G_{q/p}(x_1,\mu_f)G_{g/p}(x_2,\mu_f) + (x_1\leftrightarrow x_2)] \\
& \times &
(\hat{\sigma}^{\text{born}}
+\hat{\sigma}^{\text{virt}}
+\hat{\sigma}^{\text{soft}}
+\hat{\sigma}^{\overline{\text{HC}}})
+ d\sigma^{\text{coll}} \} \\
& + & \sum_{(\alpha,\beta)}\int dx_1dx_2 [G_{\alpha/p}(x_1,\mu_f)G_{\beta/p}(x_2,\mu_f)
+ (x_1\leftrightarrow x_2)] \hat{\sigma}^{\overline{\text{C}}}
(\alpha\beta\rightarrow q^{\ast}+X).
\end{eqnarray}
Note that the above expression contains no singularities since all the infrared divergences are canceled out.

\subsection{Virtual and real corrections for the decay of the excited quark}

Following the above procedure in the production process,
we can obtain the NLO corrections for the decay of the excited quark.
The virtual corrections for the decay part of channel $q^{\ast}\rightarrow qg$ have the same results as the production, and for the decay part of channel $q^{\ast}\rightarrow q\gamma$ the virtual corrections are
\begin{eqnarray}\label{Eq. virt decay gamma}
\nonumber
\mathcal{M}_{q^{\ast}\rightarrow q\gamma}^{\text{virt}} &=& \mathcal{M}_{q^{\ast}\rightarrow q\gamma}^{\text{born}}
  \frac{\alpha_s}{3\pi}C_{\epsilon}\bigg[
  -\frac{1}{\epsilon_{\text{IR}}^2}+\frac{1}{\epsilon_{\text{IR}}}\left(-\frac{5}{2}
  -\ln(\mu_r^2/m_{q^{\ast}}^2)\right) \\
&& -6-\frac{\pi ^2}{6}-\frac{7}{2}\ln(\mu_r^2/m_{q^{\ast}}^2)-\frac{1}{2}\ln^2(\mu_r^2/m_{q^{\ast}}^2) \bigg].
\end{eqnarray}
Furthermore,
the soft and collinear results for the decay part are given by
\begin{eqnarray}
\nonumber
d\hat{\sigma}_{q^{\ast}\rightarrow q\gamma+g}^{\text{soft}} &=& d\hat{\sigma}_{q^{\ast}\rightarrow q\gamma}^{\text{born}}
\frac{\alpha_s}{3\pi}C_{\epsilon}\bigg[\frac{2}{\epsilon_{\text{IR}}^2}
+\frac{1}{\epsilon_{\text{IR}}}\bigg(2+2\ln\frac{\mu_r^2}{m_{q^{\ast}}^2}-4\ln\delta_s\bigg) \\
&& + 4\ln^2\delta_s -4\ln\delta_s\left(\ln\frac{\mu_r^2}{m_{q^{\ast}}^2}+1\right) + \ln^2\frac{\mu_r^2}{m_{q^{\ast}}^2} + 2\ln\frac{\mu_r^2}{m_{q^{\ast}}^2} -\frac{\pi ^2}{3} +4 \bigg],
\end{eqnarray}
\begin{eqnarray}
\nonumber
d\hat{\sigma}_{q^{\ast}\rightarrow q\gamma+g}^{\text{coll}} &=& d\hat{\sigma}_{q^{\ast}\rightarrow q\gamma}^{\text{born}}
\frac{\alpha_s}{3\pi}C_{\epsilon}\bigg[
\frac{1}{\epsilon_{\text{IR}}}\left(3+4\ln\delta_s\right) \\
&& - 4\ln\delta_c\ln\delta_s -3\ln\delta_c -2\ln^2\delta_s + 4\ln\delta_s\ln\frac{\mu_r^2}{m_{q^{\ast}}^2} +3\ln\frac{\mu_r^2}{m_{q^{\ast}}^2} -\frac{2\pi^2}{3} +7 \bigg],
\end{eqnarray}
\begin{eqnarray}
d\hat{\sigma}_{q^{\ast}\rightarrow qg+g}^{\text{soft}}/d\hat{\sigma}_{q^{\ast}\rightarrow qg}^{\text{born}}
 &=& d\hat{\sigma}_{qg\rightarrow q^{\ast}+g}^{\text{soft}}/d\hat{\sigma}_{qg\rightarrow q^{\ast}}^{\text{born}},
\end{eqnarray}
\begin{eqnarray}
\nonumber
d\hat{\sigma}_{q^{\ast}\rightarrow qg+g}^{\text{coll}} &=& d\hat{\sigma}_{q^{\ast}\rightarrow qg}^{\text{born}}
\frac{\alpha_s}{3\pi}C_{\epsilon}\bigg[
\frac{1}{\epsilon_{\text{IR}}}\bigg(\frac{45}{4}-\frac{N_f}{2}+13\ln\delta_s\bigg) \\
\nonumber
&& -13\ln\delta_c\ln\delta_s -\frac{45}{4}\ln\delta_c -\frac{13}{2}\ln^2\delta_s +13\ln\delta_s\ln\frac{\mu_r^2}{m_{q^{\ast}}^2} \\
&& + \left(\frac{45}{4}-\frac{N_f}{2}\right)\ln\frac{\mu_r^2}{m_{q^{\ast}}^2} +\frac{N_f}{2}\ln\delta_c -\frac{5N_f}{6} -\frac{13\pi^2}{6} +\frac{95}{4} \bigg].
\end{eqnarray}
After adding the above real emissions and the virtual corrections, all the IR divergences can be   canceled.

\section{Numerical Discussions}
\label{sec:numerical}

In this section, we give the numerical results of the total and differential cross sections. In the decay part, we discuss the channel $q^{\ast}\rightarrow q\gamma$ and $q^{\ast}\rightarrow qg$. Only the first-generation excited quarks ($u^{\ast}$, $d^{\ast}$) would be expected to predominantly be produced in $pp$ collisions, as considered in Refs.\cite{Baur:1989kv, Bhattacharya:2009xg}, so we only present the results of $u^{\ast}$. We focus on the scenario where the compositeness scale is the same as the mass of the excited quark, i.e., $\Lambda=m_{q^{\ast}}$, and assume that $f_s$, $f$ and $f^{\prime}$ are taken as the same value: $f_s=f=f^{\prime}=1$. Other SM input parameters are:
\begin{eqnarray}
\nonumber
  && \alpha=1/128,  \quad  \alpha_s(m_Z)=0.118, \quad  \sin^2\theta_{W}=0.2312, \\
 && m_W= 80.385 \textrm{~GeV}, \quad m_Z= 91.188 \textrm{~GeV}.
\end{eqnarray}
The CT10 (CT10nlo) PDF sets and the corresponding running QCD coupling constant $\alpha_s$  are used in the LO (NLO) calculations. The factorization and renormalization scales are set as $\mu_f=\mu_r=\mu_0$, and $\mu_0=m_{q^{\ast}}$.

\begin{figure}[h]
  \centering
  \includegraphics[scale=0.4]{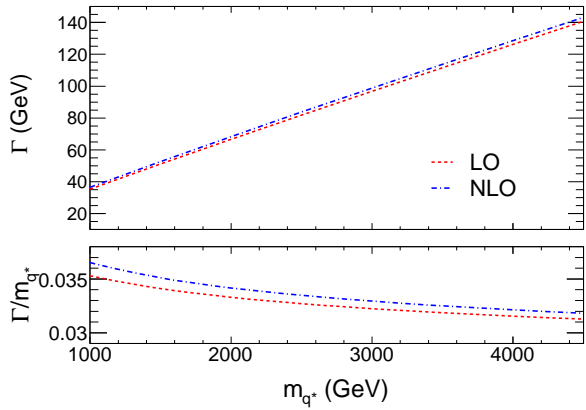} \\
  \caption{The LO and NLO total decay widths of the excited quark with different masses.}
  \label{Fig. decwidth}
\end{figure}

The complete NLO cross section for the production and decay of the excited quark can be written as
\begin{equation}\label{Eq. tot nlo}
\sigma^{\text{NLO}}(pp\rightarrow q^{\ast}\rightarrow qV_i)=(\sigma_0+\alpha_s \sigma_1)
\left(\frac{\Gamma_0^i+\alpha_s\Gamma_1^i}{\Gamma_0+\alpha_s\Gamma_1}\right),
\end{equation}
where $\sigma_0$ is the LO contribution to the excited quark production rate; $\Gamma_0$ is the LO total decay width; and $\Gamma_0^i$ is the LO decay width in the channel $q^{\ast}\rightarrow qV_i$. Here, $\sigma_1$, $\Gamma_1$ and $\Gamma_1^i$ represent their NLO corrections. We expand Eq. (\ref{Eq. tot nlo}) to order $\alpha_s$ as
\begin{equation}
\sigma^{\text{NLO}}=\sigma_0\times\frac{\Gamma_0^i}{\Gamma_0}+\alpha_s\sigma_1\times\frac{\Gamma_0^i}{\Gamma_0}
+\sigma_0\times\frac{\alpha_s\Gamma_1^i}{\Gamma_0}-\sigma_0\times\frac{\Gamma_0^i}{\Gamma_0}\frac{\alpha_s\Gamma_1}{\Gamma_0}+\mathcal{O}(\alpha_s^2).
\end{equation}
Now we turn to the calculations of the NLO QCD corrections to the total decay width of the excited quark. The analytical results of the LO decay width have been presented in Eq. (\ref{Eq. LO decay width}). We show the LO and NLO total decay widths with different excited quark masses in Fig. \ref{Fig. decwidth}.

We use the $\text{anti-}k_t$ jet algorithm \cite{Cacciari:2008gp} and FASTJET package \cite{Cacciari:2011ma} to combine the final-state partons into jets. For the $q\gamma$ decay channel, we set the jet radius $R=0.6$, and for the $qg$ decay channel, we set $R=0.5$. Following the experimental analysis in Ref. \cite{Chatrchyan:2013qha}, we consider wide jets as the final states, which are formed by clustering additional jets into the closest leading jet if within a distance $\Delta R=\sqrt{\Delta\eta^2+\Delta\phi^2}< 1.1$, and $\Delta\eta$ and $\Delta\phi$ are the pseudorapidity and azimuthal angle differences between the jet axis and the particle kinematical direction. To account for the resolution of the detectors, the energy and momentum smearing effects are applied to the final states \cite{Aad:2009wy},
\begin{eqnarray}
\nonumber
\Delta E_j/E_j &=& 0.50/\sqrt{E_j/\text{GeV}}\oplus 0.03, \\
\Delta E_{\gamma}/E_{\gamma} &=& 0.10/\sqrt{E_{\gamma}/\text{GeV}}\oplus 0.01,
\end{eqnarray}
where $E_{j,\gamma}$ are the energies of the jets and photons, respectively.
We also require the final-states particles to satisfy the following basic kinematic cuts. For the $q\gamma$ decay channel,
\begin{equation}
p_t^{\gamma}>85\text{GeV}, \quad p_t^j>30\text{GeV}, \quad \eta^{\gamma}<1.37, \quad \eta^j<2.8, \quad \Delta R_{\gamma j}>0.6;
\end{equation}
For the $qg$ decay channel,
\begin{equation}
\quad p_t^j>30\text{GeV}, \quad \eta^j<2.5,
\end{equation}
respectively. Here, $p_t^{j,\gamma}$ and $\eta^{j,\gamma}$ are the transverse momentum and pseudorapidity of the final-state jets and photon, respectively. $\Delta R_{\gamma j}$ is used to isolate the photon from jets.

\begin{figure}[h]
  \centering
  \includegraphics[scale=0.4]{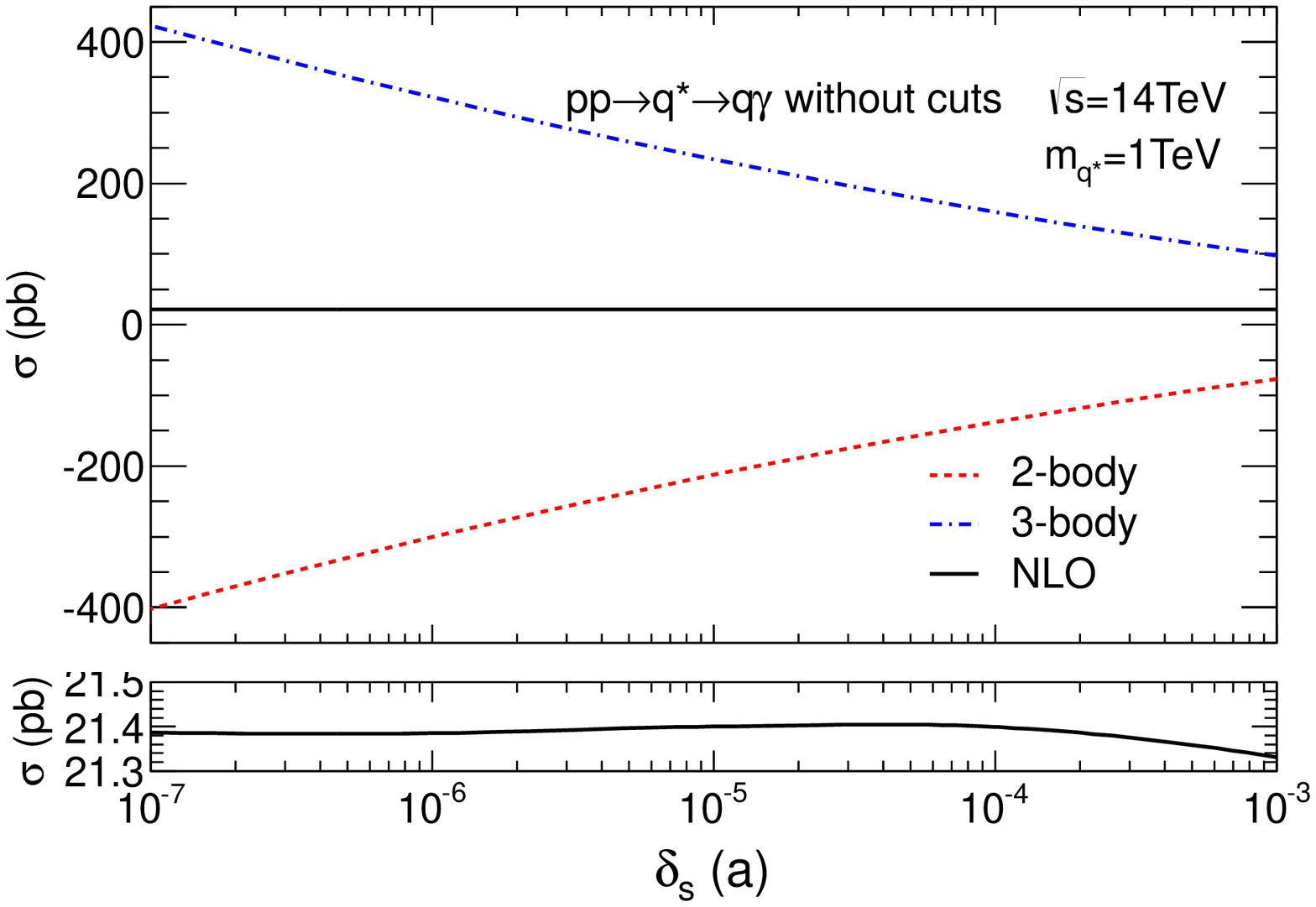} \includegraphics[scale=0.4]{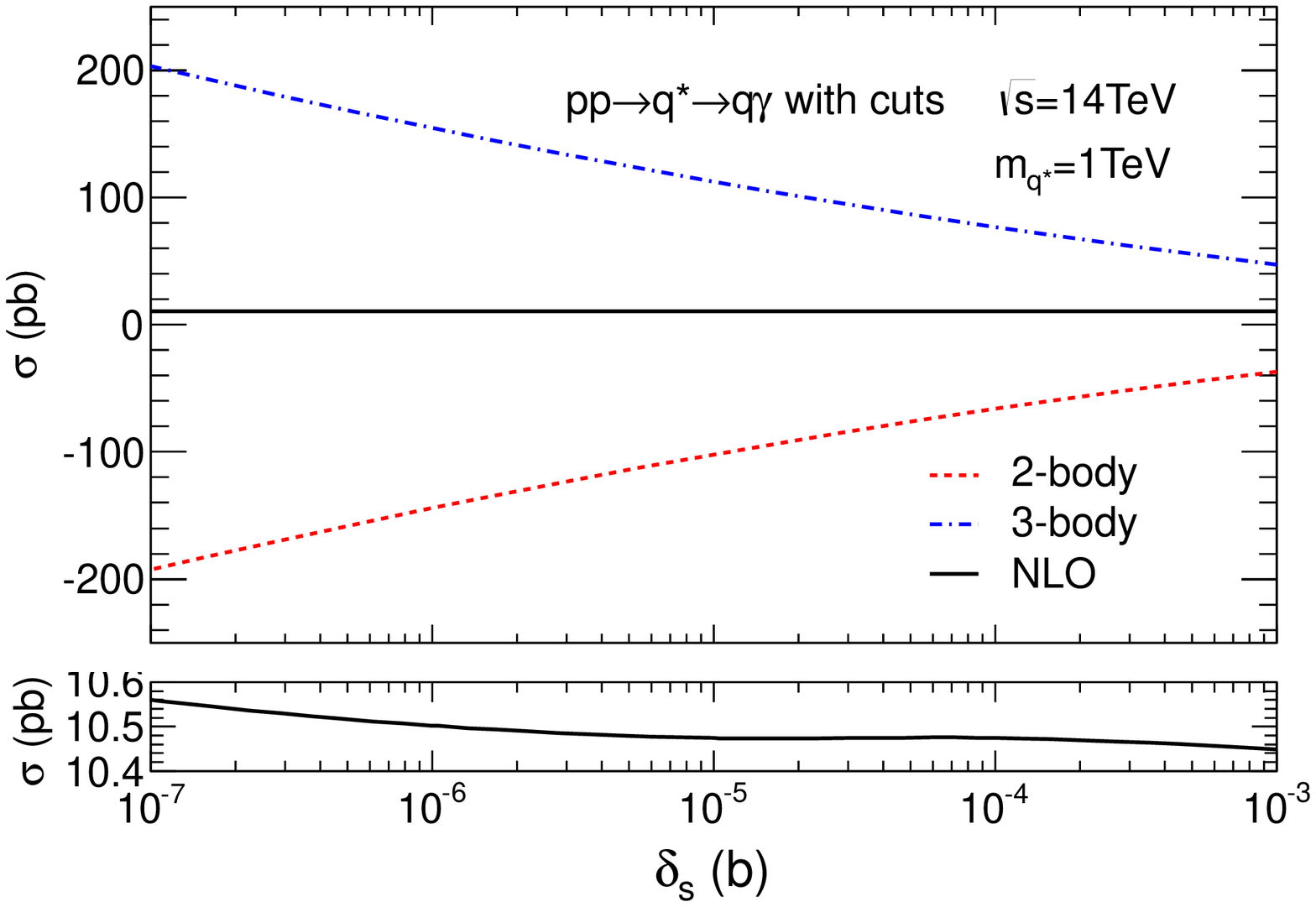} \\
  \includegraphics[scale=0.4]{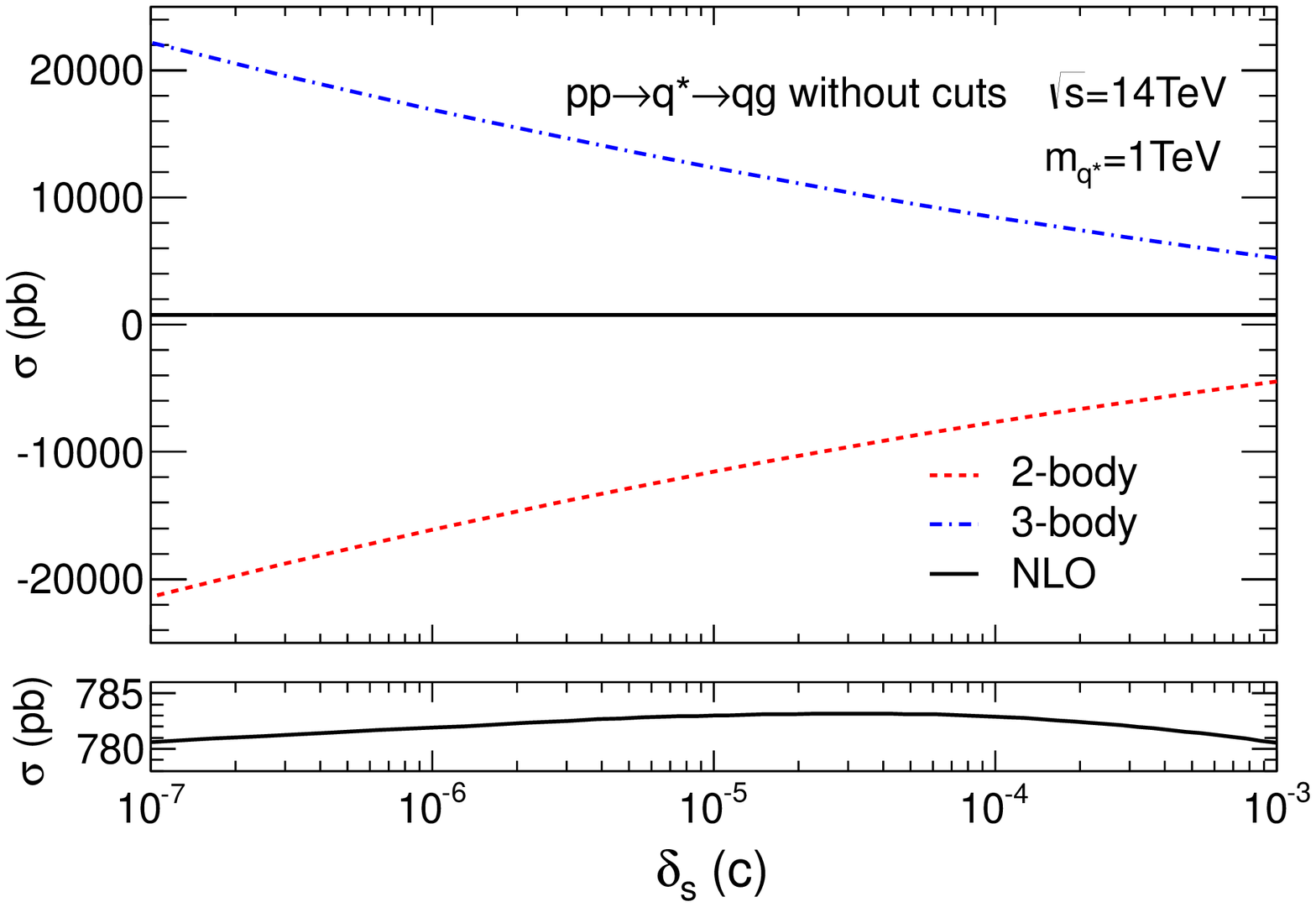} \includegraphics[scale=0.4]{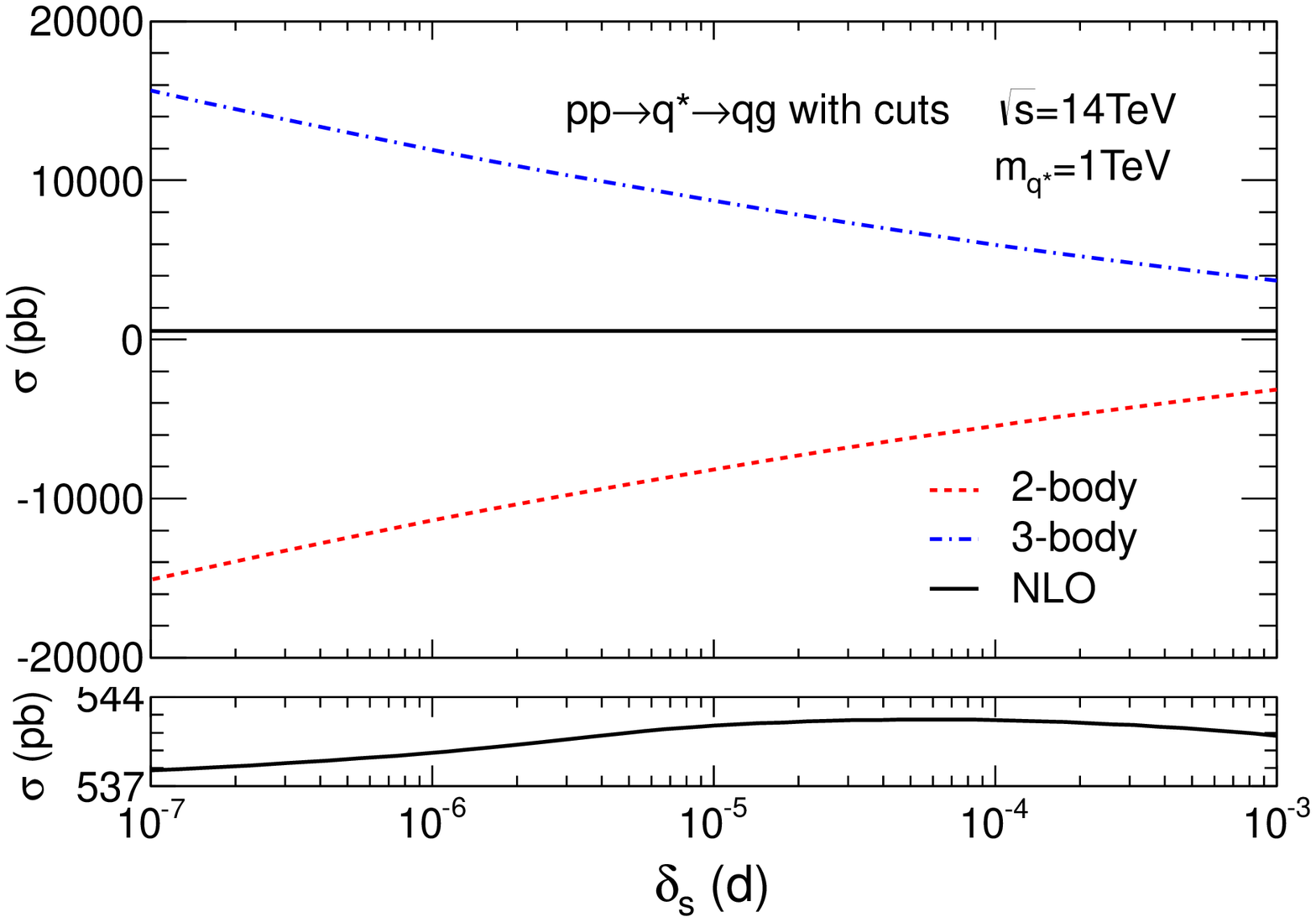} \\
  \caption{The dependence of the cross sections on the cutoffs $\delta_s$ with $\delta_c=\delta_s/50$. Panels (a) and (b) represent the $\delta_s$ dependence of the process $pp\rightarrow q^{\ast}\rightarrow q\gamma$, and (c) and (d) represent the $\delta_s$ dependence of the process $pp\rightarrow q^{\ast}\rightarrow qg$. Panels (a) and (c) are calculated before any cut is imposed, and (b) and (d) are calculated after all basic and additional cuts are imposed.  The additional cuts are discussed in Sec. \ref{sec:simulation}.}
  \label{Fig. deltas}
\end{figure}

In Fig.~\ref{Fig. deltas} we show the dependence of the NLO cross sections on the cutoffs $\delta_s$ and $\delta_c$. From Fig.~\ref{Fig. deltas} we can see that the soft-collinear and the hard-noncollinear parts individually  strongly depend on the cutoffs, but the total cross section is independent of the cutoffs. Figure \ref{Fig. deltas} also show that $\sigma^{\text{NLO}}$ only changes slightly with $\delta_s$ varying from  $10^{-7}$ to $10^{-3}$, which indicates that it is reasonable to use the two cutoff phase-space slicing method. In the following calculations, we take $\delta_s=10^{-4}$ and $\delta_c=\delta_s/50$.

\begin{figure}[h]
  \centering
  \includegraphics[scale=0.4]{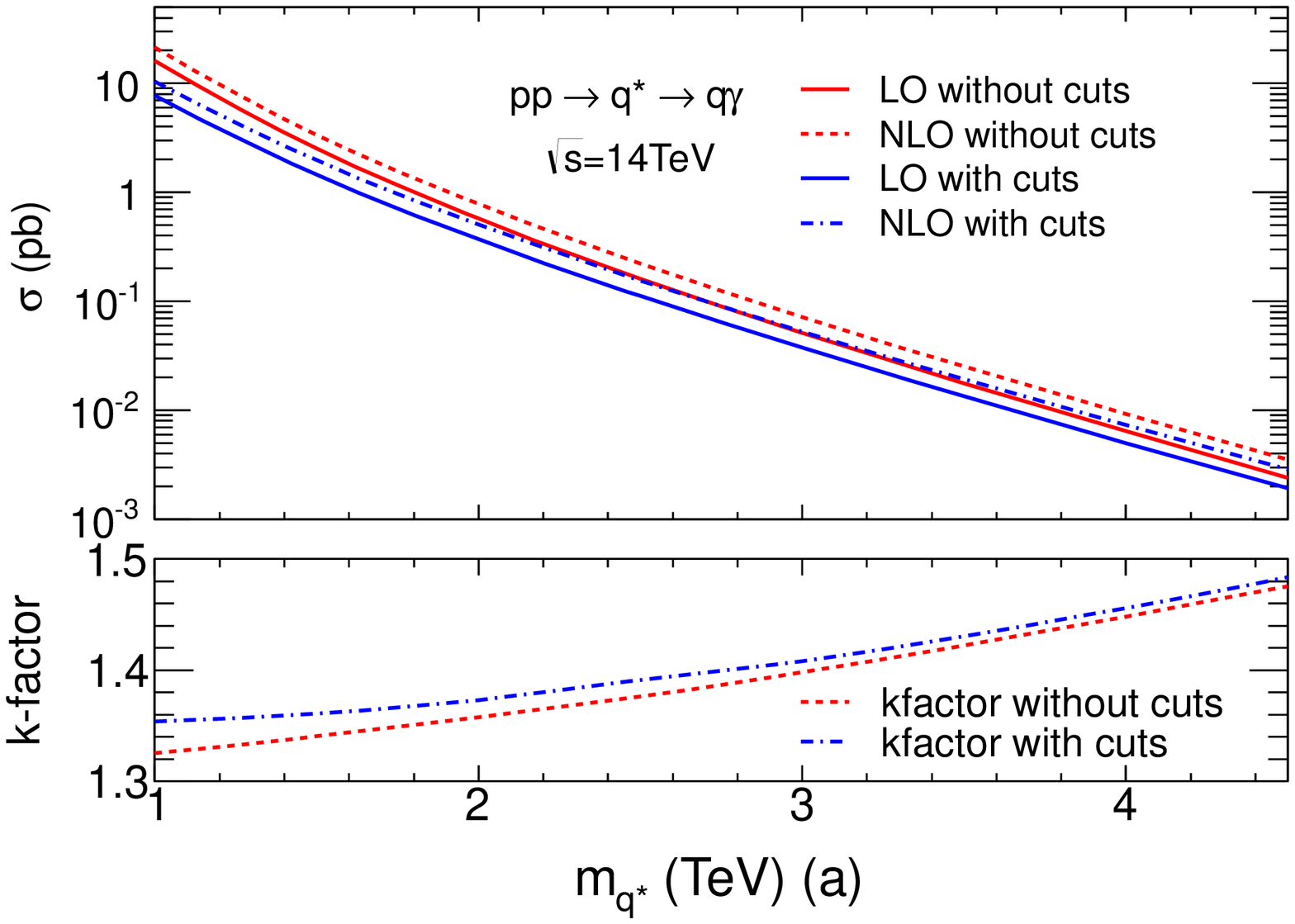} \includegraphics[scale=0.4]{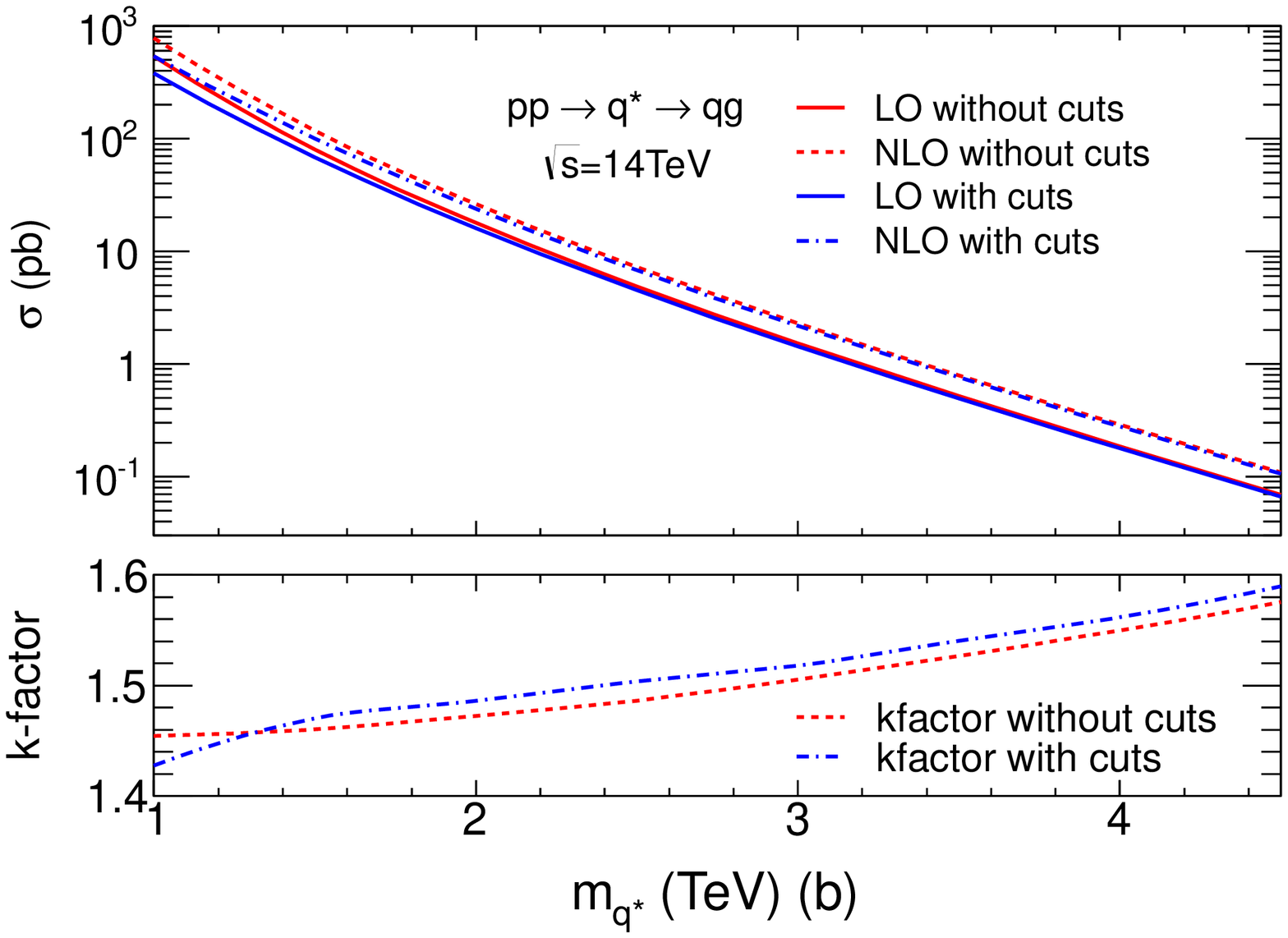} \\
  \caption{The total cross section and the k factors as functions of the excited quark mass. Panel (a) represents the $q\gamma$ decay channel, and (b) represents the $qg$ decay channel. The lines labeled as "with cuts" represent the results after all basic and additional kinematical cuts have been imposed.}
  \label{Fig. kfactor}
\end{figure}

Figure \ref{Fig. kfactor} plots the total cross section and the k factor defined as $\sigma^{\text{NLO}}/\sigma^{\text{LO}}$, as the function of the excited quark mass, which shows that the QCD NLO corrections are more significant for larger excited quark mass. Here, compared to the $q\gamma$ decay channel, the k factors for the $qg$ decay channel are relatively larger. And in general, the k factors are slightly enhanced after all cuts are  imposed. From Fig. \ref{Fig. kfactor}, we can see that the k factor varies from about $30\%$ to $50\%$ and from about $40\%$ to $60\%$ for different excited quark masses in the $q\gamma$ and $qg$ decay channels, respectively.

\begin{figure}[h]
  \centering
  \includegraphics[scale=0.40]{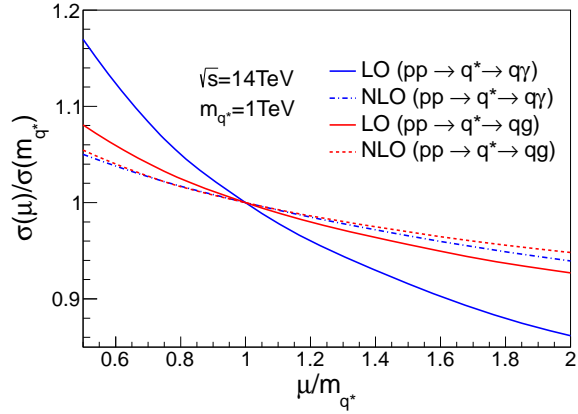} \\
  \caption{The scale dependence of the LO and NLO cross sections with $q\gamma$ and $qg$ decay modes, respectively. We set the factorization scale and renormalization scale as $\mu_f=\mu_s=\mu$.}
  \label{Fig. scale dependence}
\end{figure}

Figure~\ref{Fig. scale dependence} gives the scale dependencies of the LO and NLO total cross sections for $q\gamma$ and $qg$ decay channels, which shows that the NLO corrections significantly reduce the scale dependencies for both decay channels, and make the theoretical predictions more reliable.

\begin{figure}[h]
  \centering
   \includegraphics[scale=0.40]{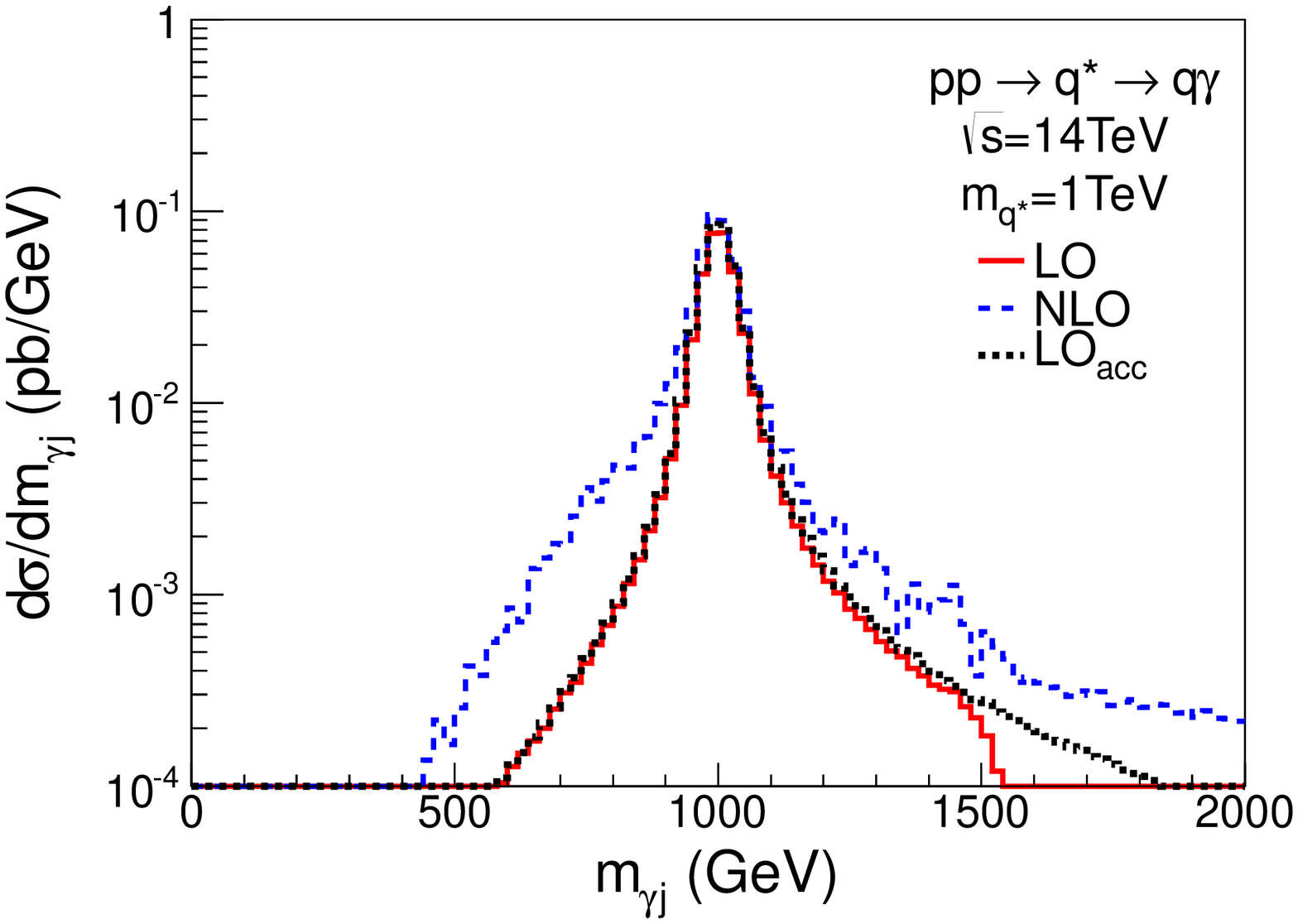} \includegraphics[scale=0.40]{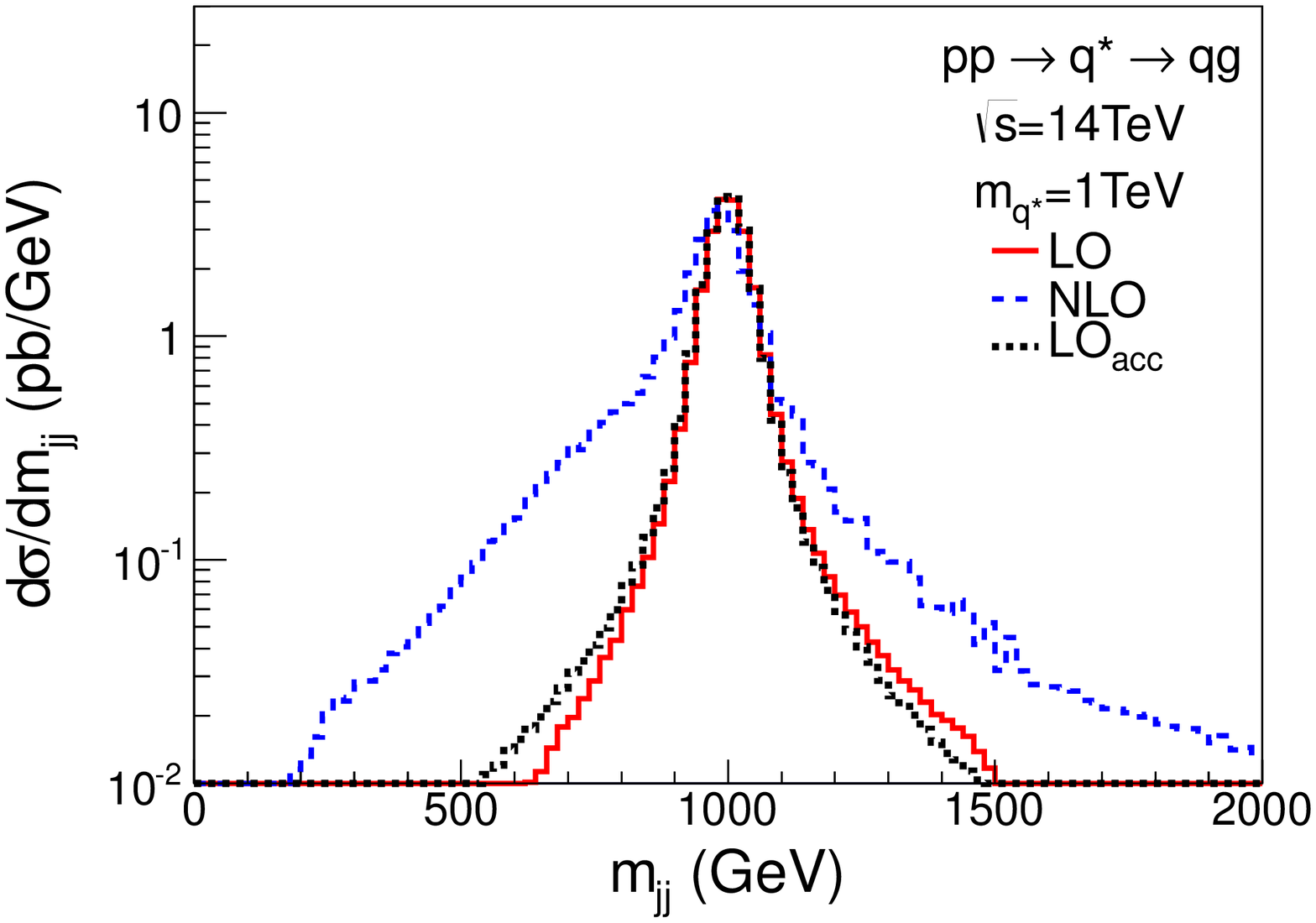}\\
  \caption{Distributions of the invariant mass of the final-state photon+jet and dijet.  Here, LO and NLO represent the results in the narrow width approximation, and LO$_{\text{acc}}$ represents the full LO result. }
  \label{Fig. Minv}
\end{figure}

\begin{figure}[h]
  \centering
   \includegraphics[scale=0.40]{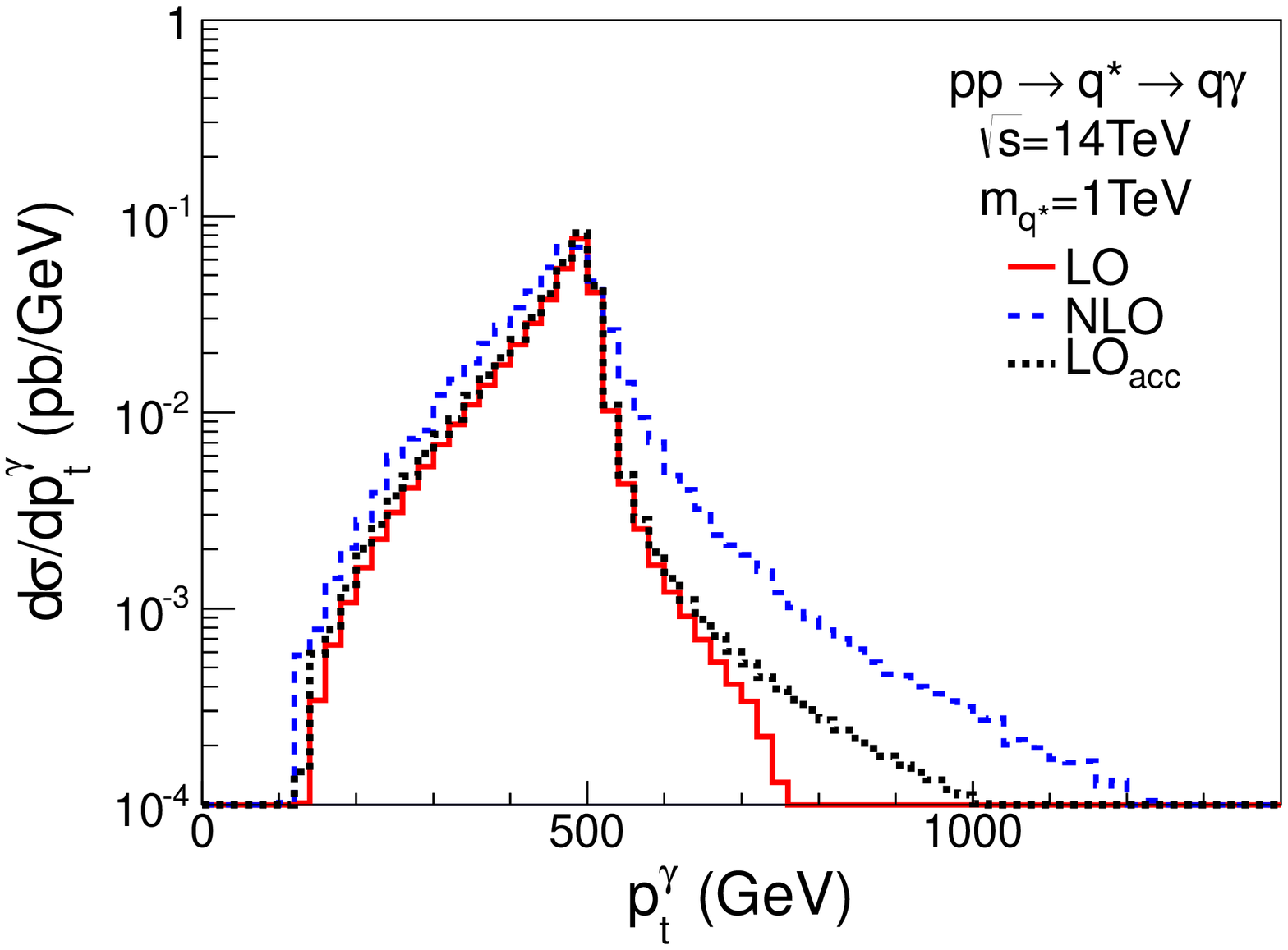} \includegraphics[scale=0.40]{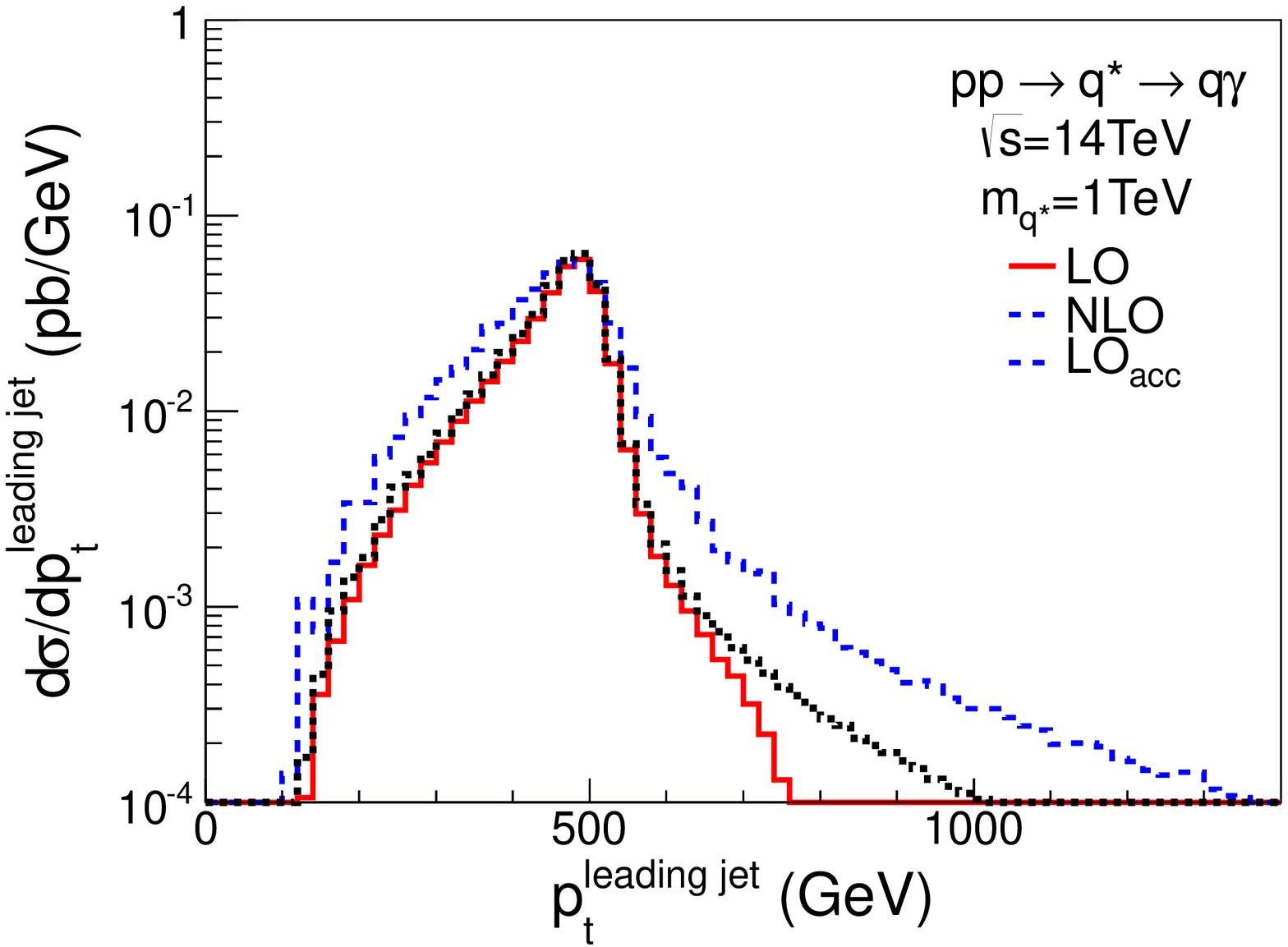}\\
   \includegraphics[scale=0.40]{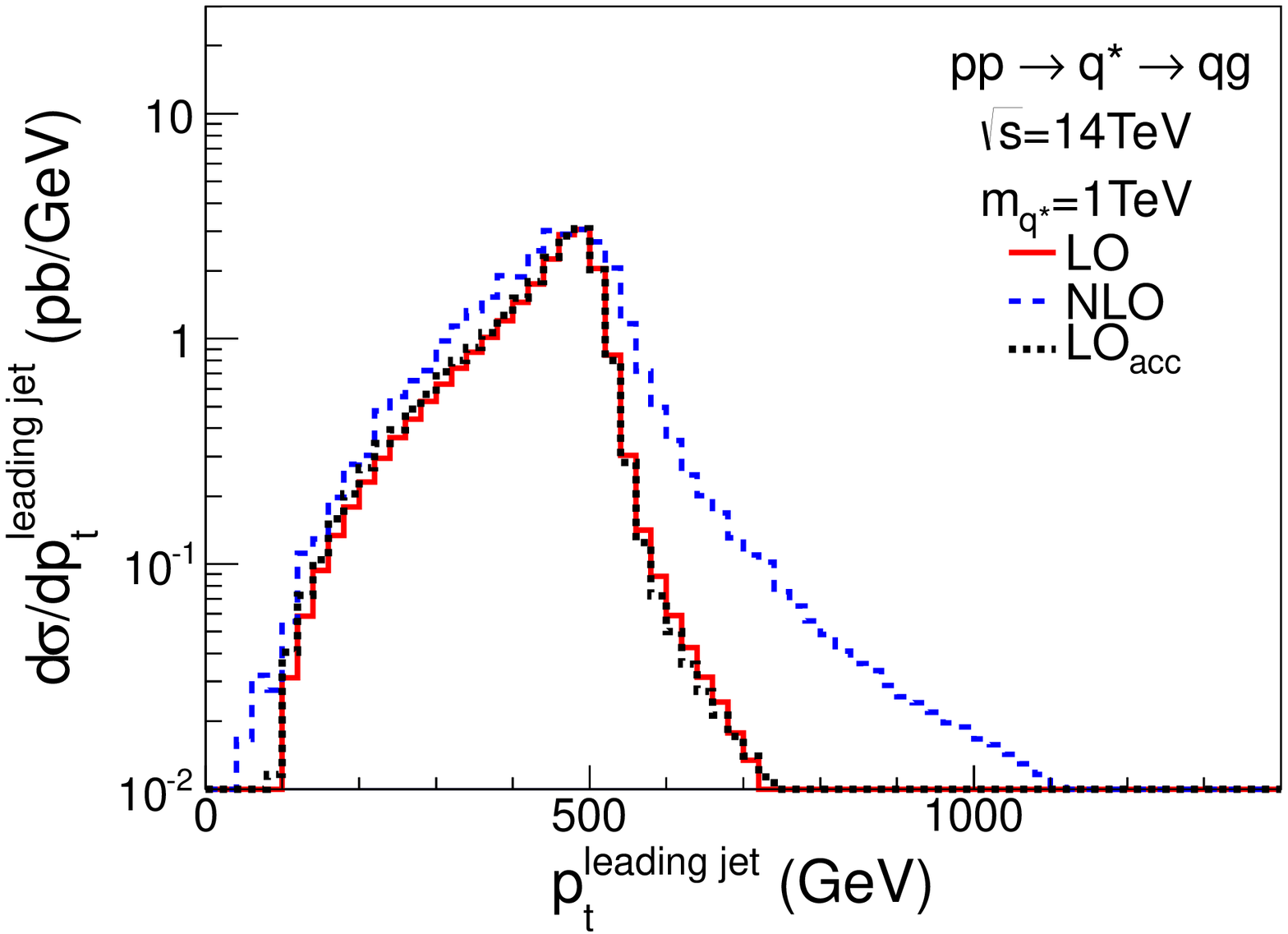} \includegraphics[scale=0.40]{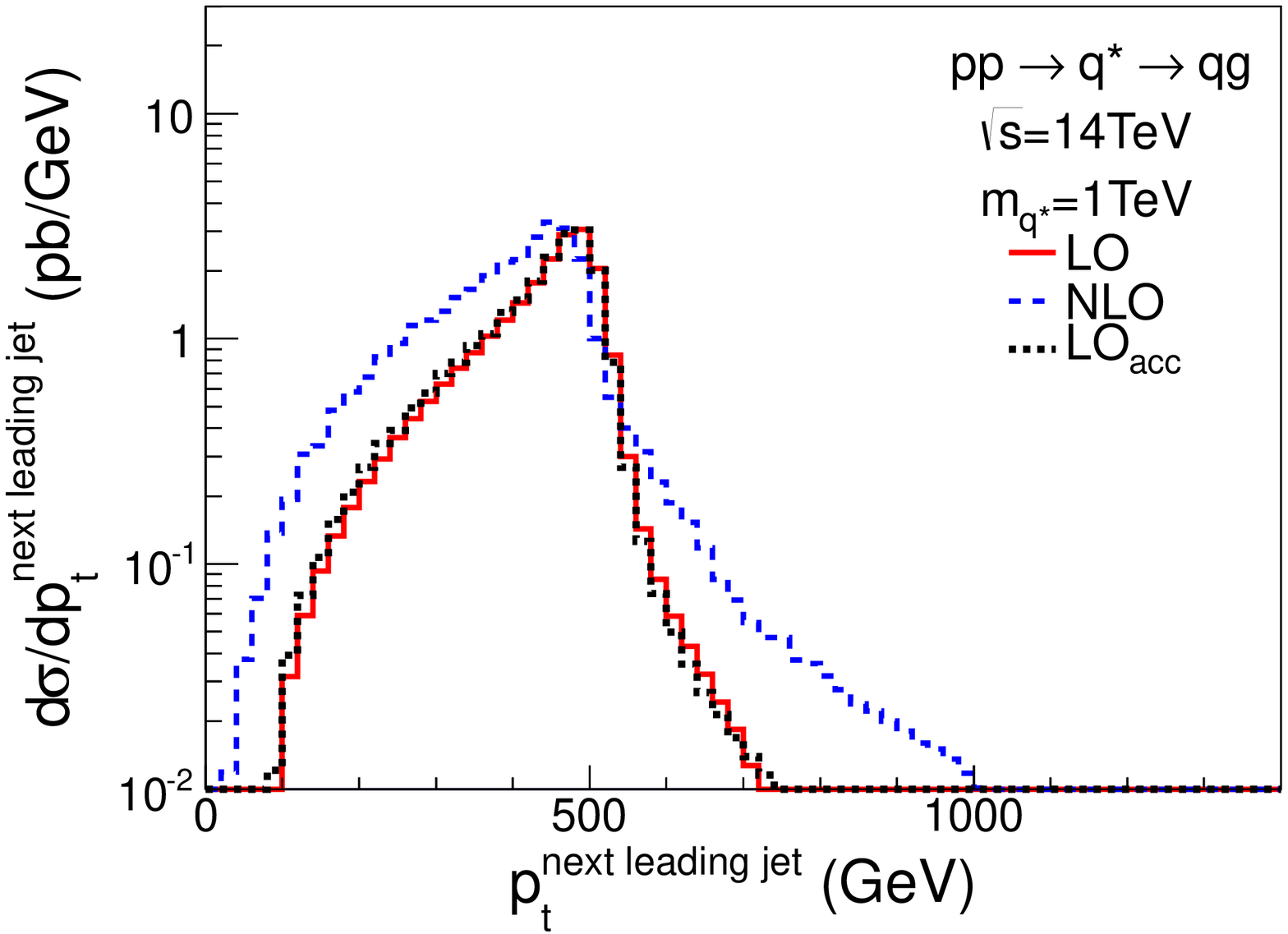}\\
  \caption{The LO and NLO differential cross sections in the transverse momentum $p_t$ of the photon and leading jet in  $\gamma$+jet production and the leading and next-to-leading jets in  dijet production. Here, LO and NLO represent the results in the narrow width approximation, and LO$_{\text{acc}}$ represents the full LO result. }
  \label{Fig. pt}
\end{figure}

Figure \ref{Fig. Minv} shows the invariant mass distributions with the $q\gamma$ and $qg$ decay channels. We can see that the distributions have sharp peaks around the excited quark mass, and the NLO corrections broaden the distributions  but do not change the position of the peaks.

In Fig. \ref{Fig. pt}, we display differential cross sections for the transverse momentum $p_t$ of the photon and jet. We find that the NLO QCD corrections enhance the LO results at both low $p_t$ and high $p_t$. There is a sharp fall in the $p_t$ distribution at about half the excited quark mass, which is called the Jacobian edge \cite{Gordon:1998bn}. The edge is broadened by the excited quark width and real corrections at NLO.

In Figs.\ref{Fig. Minv} and \ref{Fig. pt}, we also compare the LO differential distributions in the narrow width approximation with the ones without such an approximation. There are only slight changes in the shapes. Then, it can be seen that using the narrow width approximation is reasonable in our scenario.

\section{Signal and Background}
\label{sec:simulation}

The dominant backgrounds for the process $pp\rightarrow q^{\ast}\rightarrow q\gamma$ are $qg\rightarrow q\gamma$ and $q\bar{q}\rightarrow g\gamma$ at the leading order in the SM, which are irreducible backgrounds. The second-largest background comes from the QCD dijet production, where one of the jets mimics an isolated photon. The production of $W/Z+\gamma$ also yields similar final states, but owing to their negligible contributions \cite{Khachatryan:2014aka}, it is not considered in this work. Moreover, the main background for the process $pp \rightarrow q^{\ast}\rightarrow qg$ comes from the dijet production in the SM. Another background comes from the production of $W$ and $Z$ bosons (with their subsequent decay into jets), but it makes small contributions compared to the QCD jets background \cite{Cakir:2000vt} so it is also neglected in this discussion.

To optimize the expected significance of signals, we require the following additional cuts on the final states. For the $q\gamma$ decay channel,
\begin{equation}
p_t^{\gamma}>170\text{GeV}, \quad p_t^j>170\text{GeV}, \quad m_{\gamma j}>890\text{GeV}.
\end{equation}
For the $qg$ decay channel,
\begin{equation}
H_T>650\text{GeV}, \quad m_{\gamma j}>890\text{GeV},
\end{equation}
respectively. Here, $H_T$ denotes the scalar sum of the jet $p_t$. We use the
MadGraph5$\_$aMC@NLO
package \cite{Alwall:2011uj} to calculate the corresponding
LO and NLO
backgrounds, respectively.
The probability $P_{\gamma/j}$, faking a jet as a photon, is set as $P_{\gamma/j}=10^{-4}$ \cite{Baur:1992cd}.

\begin{figure}[h]
  \centering
  \includegraphics[scale=0.40]{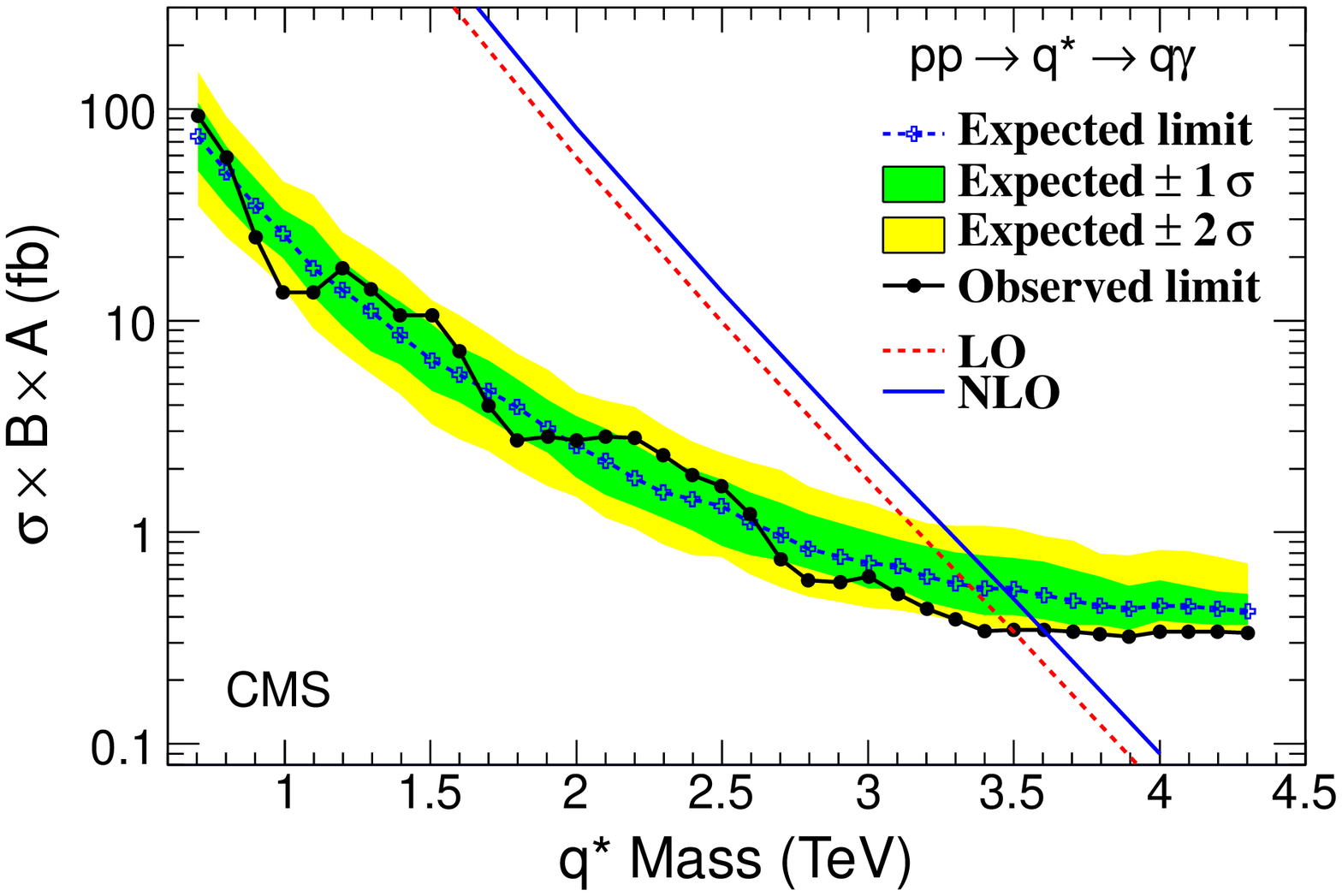}  \includegraphics[scale=0.40]{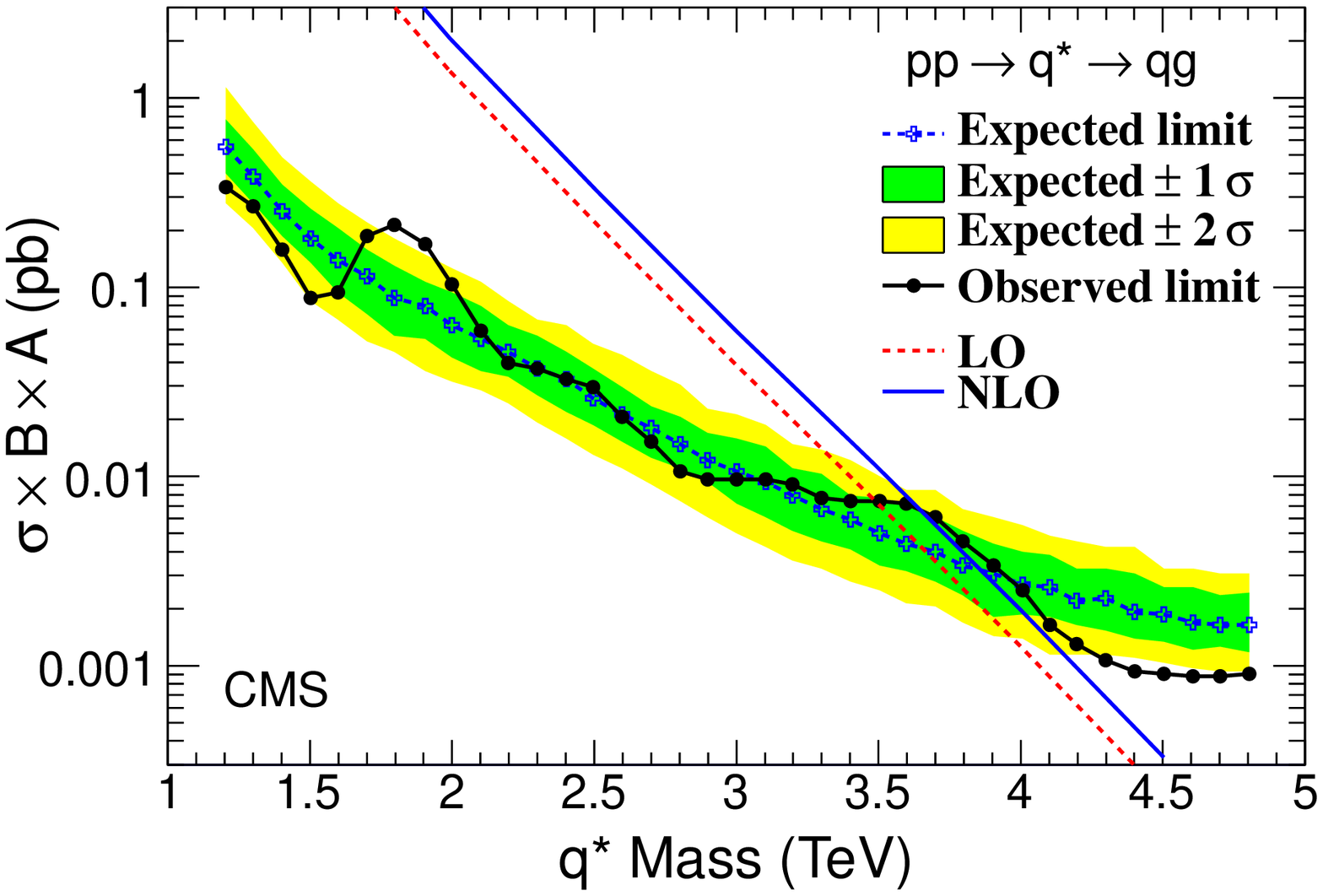}\\
  \includegraphics[scale=0.40]{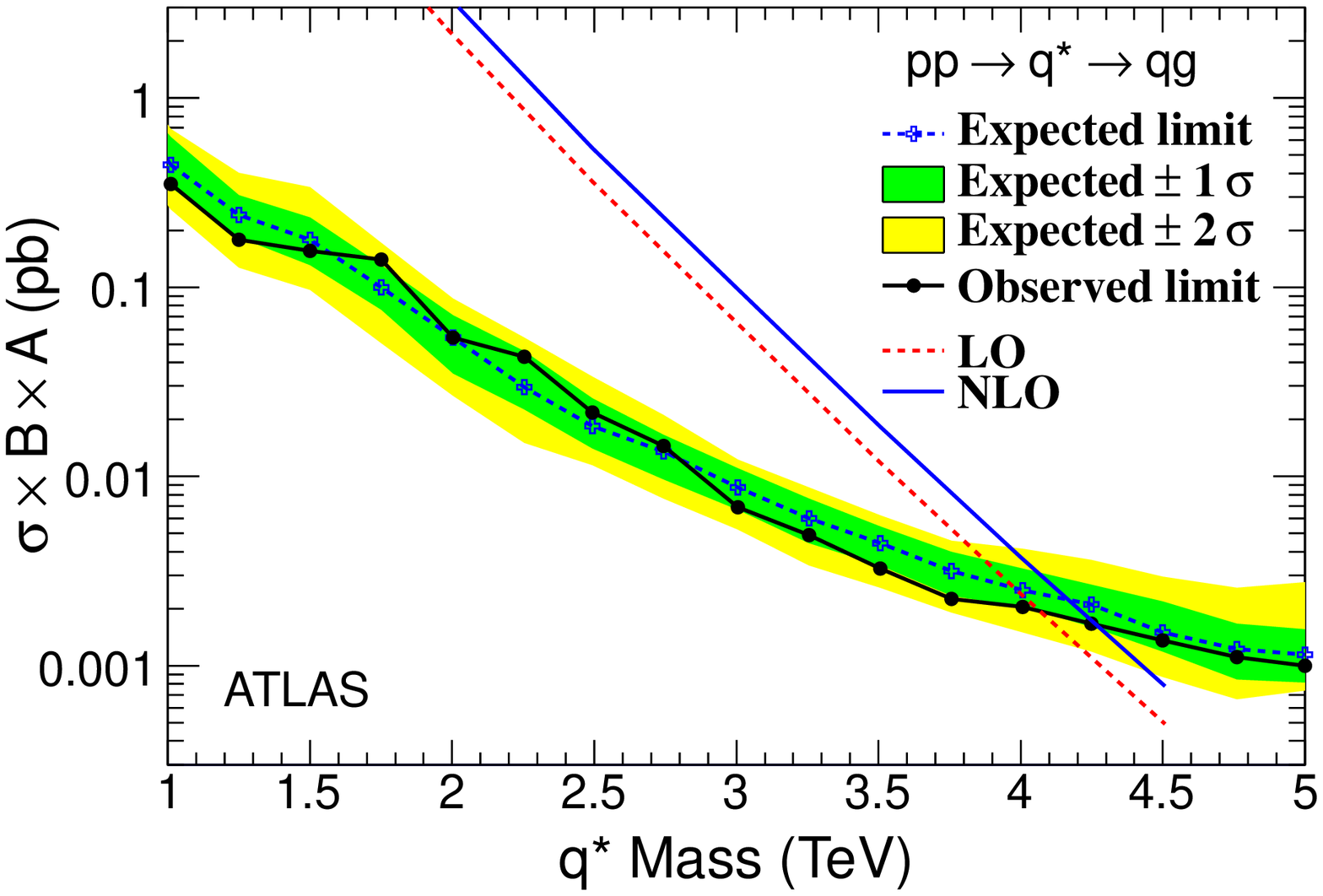}\\
  \caption{The expected and observed $95\%$ C.L. upper limits on $\sigma\times B\times A$ for the production of excited quarks in the $\gamma$+jet and dijet final states. The limits are compared with theoretical predictions. The uncertainty in the expected limits at the $1\sigma$ and $2\sigma$ levels are shown as shaded bands.}
  \label{Fig. brazil}
\end{figure}

Following Refs. \cite{Khachatryan:2014aka, Khachatryan:2015sja, Aad:2014aqa}, in Fig. \ref{Fig. brazil} we present the generic upper limits at the $95\%$ confidence level for the cross section $\sigma\times B\times A$, where $B$ and $A$ represent the branching fraction into photon+jet or dijet final states and the acceptance of the event selection by imposing the kinematic selection criteria, respectively. Due to the  QCD NLO corrections, the upper limits of the excluded mass range of the excited quark
at CMS and ATLAS
can be  promoted from 3.5 TeV to over 3.6 TeV
and from 4.1 TeV to 4.3 TeV, respectively.

\begin{figure}[h]
  \centering
   \includegraphics[scale=0.40]{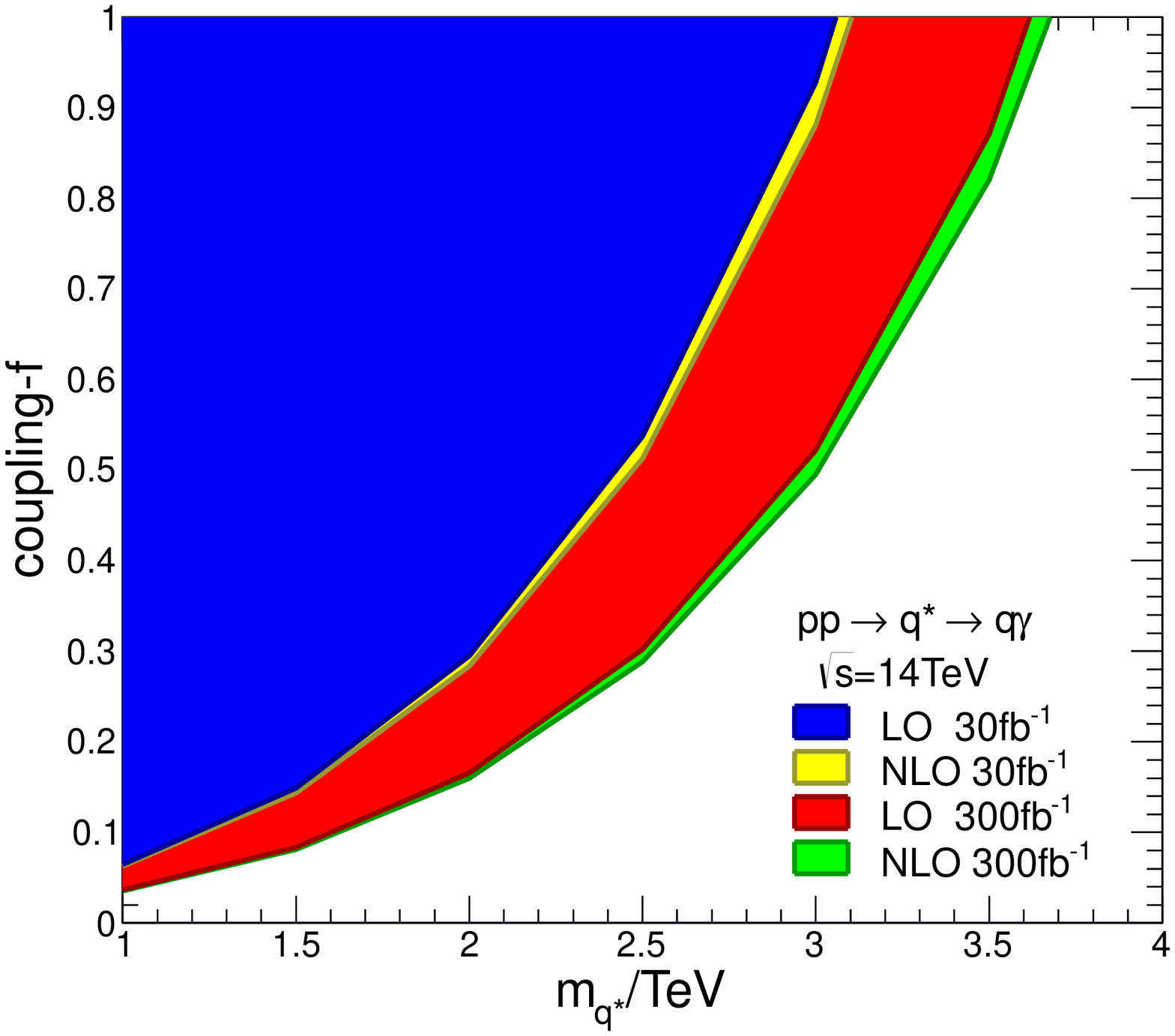} \includegraphics[scale=0.40]{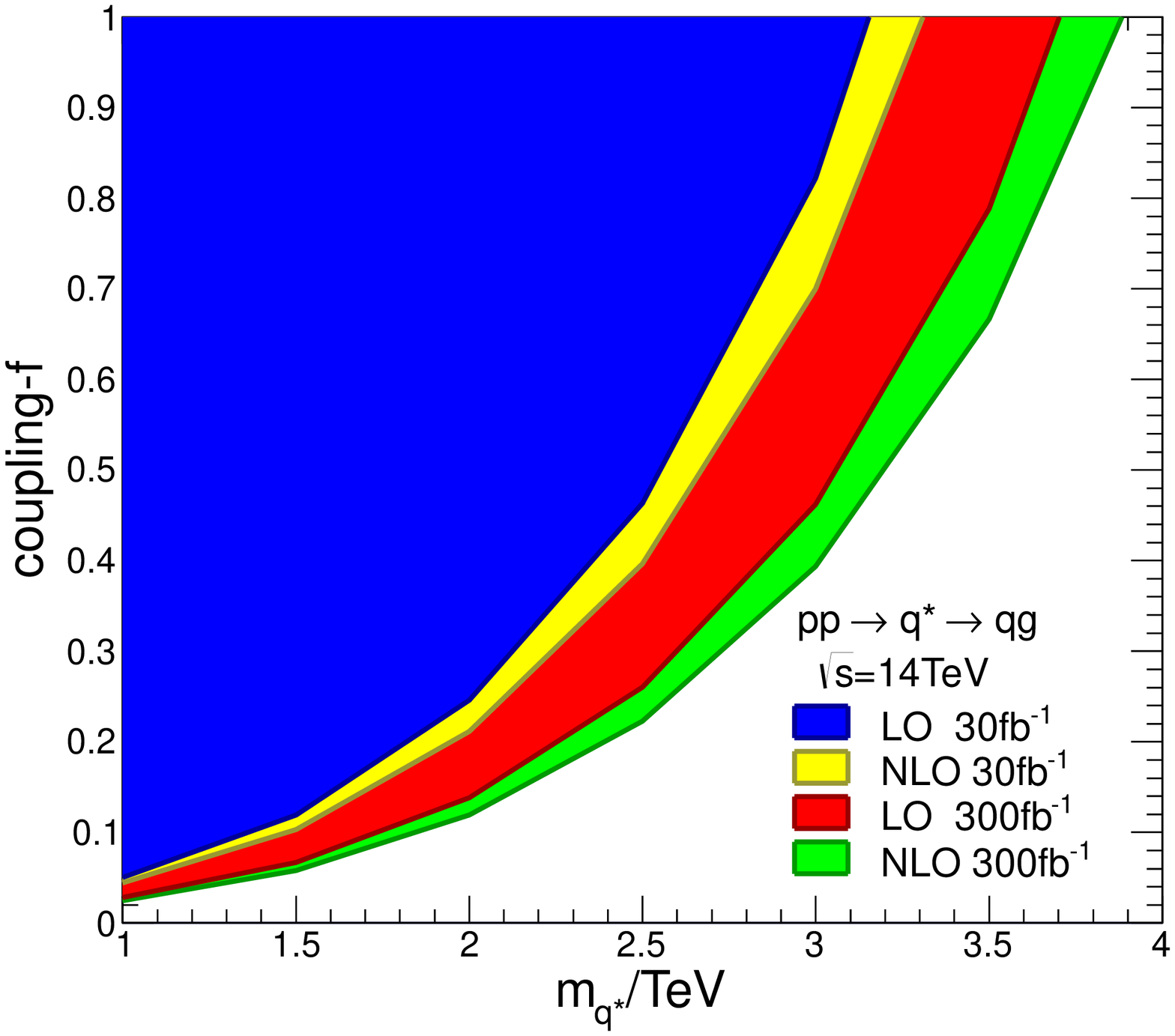}\\
  \caption{Constraints on the coupling $f$ between the excited quark and its SM partner with different excited quark masses at the 14 TeV LHC. The shaded parts in the figures are excluded regions.}
  \label{Fig. constraint}
\end{figure}

In principle, if no signal is observed, the couplings between the excited quarks and their SM partners cannot be too large. In Fig. \ref{Fig. constraint}, we show the $3\sigma$ exclusion limits ($\mathcal{S}/\sqrt{\mathcal{B}}=3$) of the couplings with several excited quark masses for integrated luminosities of 30 fb$^{-1}$ and 300 fb$^{-1}$ at the 14 TeV LHC. From Fig. \ref{Fig. constraint}, it can be seen that the NLO corrections significantly tighten the allowed parameter space
of the $q^{\ast}\rightarrow qg$ decay channel. And the parameter space of the $q^{\ast}\rightarrow q\gamma$ decay channel is slightly tightened due to large QCD NLO corrections for the background.

\section{Conclusion}
\label{sec:conclusion}

In conclusion, we have investigated the signal of the excited quark in the $\gamma$+jet and dijet final states at the LHC, including QCD NLO corrections to the production and decay of the excited quark. Our results show that the NLO corrections vary from about $30\%$ to $60\%$ for different excited quark masses, which reduce the scale dependencies of the total cross sections and tighten the constraints on the model parameters. Furthermore, we also explore the distributions for final-state invariant mass and transverse momentum with QCD NLO accuracy. Finally, we  discuss the constraints on the excited quark mass and the couplings between the excited quark and its SM partner, and find that the upper limits of the excited quark excluded mass range
at CMS and ATLAS
are promoted from 3.5 TeV to over 3.6 TeV
and from 4.1 TeV to 4.3 TeV, respectively,
and the allowed parameter space is tightened, due to the QCD NLO effects.

\begin{acknowledgments}
We would like to thank Ding Yu Shao for helpful discussions. This work was supported in part by the National Nature Science Foundation of China,
under Grants No. 11375013 and No. 11135003.
\end{acknowledgments}



\bibliography{qstar}{}

\end{document}